\documentclass[rotating]{aa}

\usepackage[final]{epsfig} 
\usepackage{epsf}
\usepackage[final]{graphics}
\usepackage{natbib}
\usepackage{amssymb}
\usepackage{color}
\usepackage{times}
\usepackage[american,english]{babel}
\begin{document}

%\headnote{Research Note}
\title{Determination of confusion noise for far-infrared measurements
%Sky confusion noise in the far-infrared II.
  \thanks{Based on observations with ISO, an ESA project with instruments
    funded by ESA Member States (especially the PI countries: France,
    Germany, the Netherlands and the United Kingdom) and
    with the participation of ISAS and NASA.}
  }
%\subtitle{Source confusion and special measurement configurations}
%\subtitle{One should find a better title for this...}
\author{ Cs.~Kiss\inst{1,2}
  %\and  P.~\'Abrah\'am\inst{1,2} 
  \and  U.~Klaas\inst{1}
  \and  D.~Lemke\inst{1}  
  %\and  Ph.~H\'eraudeau\inst{1}
  %\and  C.~del~Burgo\inst{1}
 % \and  U.~Herbstmeier\inst{1} 
  }
%\authorrunning{Cs. Kiss}
%\titlerunning{Fluctuations of the CFIRB and the small-scale structure
%   of the cirrus emission in the Galaxy}
\institute{ Max-Planck-Institut f\"ur Astronomie, K\"onigstuhl~17,
     D-69117~Heidelberg, Germany
  \and  Konkoly Observatory of the Hungarian Academy of Sciences, 
    P.O. Box 67, H-1525~Budapest, Hungary
  }
\offprints{Cs.~Kiss, pkisscs@mpia.de}
\date{ Received  / Accepted ...}
%
%%%%%%%%%%%%%%%%%%%%%%%%%%%%%%%%%%%%%%%%%%%%%%%%%%%%%%%%%%%%%%%%%%
\abstract{We present a detailed
assessment of the far-infrared confusion noise 
imposed on measurements with the ISOPHOT far-infrared 
detectors and cameras aboard the ISO satellite. We provide
confusion noise values for all measurement configurations 
and observing modes of ISOPHOT in the 
90\,$\mu$m\,$\le$\,$\lambda$\,$\le$\,200\,$\mu$m wavelength range. 
Based on these results we also give estimates for  
cirrus confusion noise levels at the resolution limits
of current and future instruments of infrared space telescopes:
Spitzer/MIPS, ASTRO-F/FIS and Herschel/PACS.  
\keywords{methods:\ observational  -- ISM:\ structure -- 
          Infrared:\ ISM:\ continuum -- diffuse radiation}}
\maketitle
%%%%%%%%%%%%%%%%%%%%%%%%%%%%%%%%%%%%%%%%%%%%%%%%%%%%%%%%%%%%%%%%%%
\section{Introduction}

Confusion noise is a major limitation 
in sensitivity and photometric accuracy
for the measurements performed 
with the far-infrared (FIR) filters/detectors of the ISOPHOT instrument
\citep{Lemke96}, on-board the Infrared Space Observatory \citep[ISO,][]{Kessler96}. 
As was shown by 
\citet[][hereafter Paper~I]{Kiss2001} measurements in 
the long-wavelength filters 
of the C100 camera (90 and 100\,$\mu$m) were affected roughly equally  
by confusion and instrument noise and measurements with the C200 detector 
were confusion noise limited. 
In Paper~I we provided estimates of the 
sky confusion noise -- the sum of cirrus confusion noise and fluctuations
of the cosmic far-infrared background -- for a special measurement 
configuration (one target position bracketed by two 
reference positions, single pixel apertures of the ISOPHOT 
C100 and C200 cameras) and for four ISOPHOT filters. 
However, it is desirable to investigate the dependence of the 
confusion noise on all actual measurement 
configurations. This is a fundamental aspect in 
the scientific validation and interpretation of FIR ISOPHOT measurements. 

Confusion noise predictions for future/current space missions working in the
far-infrared usually consider the fluctuations due to the cosmic 
far-infrared background (CFIRB) only 
%\citep[see e.g.][for Spitzer/MIPS, ASTRO-F/FIS and Herschel/PACS, respectively]{Dole,Jeong,Negrello}. 
%%%%
{  
(see e.g. Dole et al., 2003, for Spitzer/MIPS, Jeong et al., 2003, for ASTRO-F/FIS
and Negrello et al., 2004, for Spitzer/MIPS).
For deep cosmological surveys the cirrus contribution can be minimized by a 
careful selection of fields with low Galactic emission. However, most of the
FIR sky is heavily affected by this phenomenon.}
%%%%

The strength of the cirrus confusion noise is believed to decrease 
rapidly with improving spatial resolution { 
\citep[see e.g.][]{Gautier,MD2002,MD2003,Ingalls}}. 
%%%
{  
%However, since the cirrus confusion noise strongly depends on the 
%resolving power of the telescope \citep{Helou}, 
%one needs a low $\lambda$/D ratio (where $\lambda$ is the wavelength of the
%observation and D is the diameter of the telescope's primary mirror) 
%to overcome most of the cirrus contributions. 
According to the formula given by \citet{Helou}, the cirrus confusion noise
scales as ($\lambda/D$)$^{2.5}$ with $\lambda$ being the wavelength of the observation
and $D$ the diameter of the telescope primary mirror.  
In this respect the 3.5\,m Herschel
Space Telescope will be superior to other cyrogenic space missions like 
ISO, ASTRO-F or Spitzer with primary mirror diameters D\,$<$\,1m.}
%%%%
%However, only the Herschel Space 
%Telescope will have the resolving power to overcome most of the 
%cirrus noise contributions. 
Although the structure of the Galactic cirrus may 
change below the ISOPHOT resolution limit, our detailed confusion noise study 
offers the possibility to make predictions for other FIR space telescopes,
for the first time based on observations in the 170\,$\mu$m range. 
        
In this paper we present a detailed analysis of the confusion noise
for ISOPHOT measurements performed with the P3, C100 and C200 detectors
in various measurement configurations 
offered by the ISOPHOT Astronomical Observation Templates 
\citep[see the ISOPHOT Handbook,][for an overview]{Laureijs2003}.
Based on these results we 
provide predictions for the achievable photometric accuracy 
(cirrus confusion noise at the resolution limit) for 
the Spitzer/MIPS, ASTRO-F/FIS and Herschel/PACS instruments.  	
%%%%%%%%%%%%%%%%%%%%%%%%%%%%%%%%%%%%%%%%%%%%%%%%%%%%%%%%%%%%%%%%%%%

\section{ISOPHOT instrumental set-up and observational parameters}
%\section{Confusion noise ISOPHOT measurements}

\subsection{Apertures, filters and detector arrays}

Since the strength of the confusion noise is highly wavelength
dependent (see Paper~I), we considered only those ISOPHOT filters,
where the confusion noise is at least as strong as the typical
value of the instrument noise. These are the filters with central
wavelengths $\lambda$\,$\ge$\,90\,$\mu$m. Some 
filters lack the required number of appropriate maps, therefore we
restricted our analysis to the following detector/filter combinations:
C100: 90 and 100\,$\mu$m, C200: 170 and 200\,$\mu$m 
\citep[see the ISOPHOT Handbook,][]{Laureijs2003}. 
Confusion noise analysis has been performed for single pixels and
for the whole detector array field-of-view as well. 
This is necessary, because for some measurement modes the photometric
flux is derived from the summed-up fluxes of the array,
e.g. for C200 staring, where the source is centered on the common
corner of the four detector pixels. 
In the case of single pixels the size 
of the target/reference apertures are equal to 
46\arcsec$\times$46\arcsec and 
92\arcsec$\times$92\arcsec for the C100 and C200 camera,
respectively. In the case of the full arrays the size of the 
target/reference apertures are equal to 
138\arcsec$\times$138\arcsec~[3$\times$3 array] and 
184\arcsec$\times$184\arcsec~[2$\times$2 array], respectively. 
Although there were no suitable "maps" for the P3 detector -- 
100\,$\mu$m filter combination, this was modeled
with the help of maps
obtained by the C100 camera in its 100\,$\mu$m filter. 
The system responses of both filters are quite similar
\citep[see][Appendix~A]{Laureijs2003}.
Five model apertures were constructed using the C100 detector
pixel granulation, and corresponding to the 79\arcsec, 99\arcsec, 
120\arcsec\, and 180\arcsec\, circular and to the 
127\arcsec$\times$127\arcsec\, rectangular apertures.
These model "apertures" were generated by 
6$\times$6 pixel matrices on the
C100 maps, with weights for each pixel suitably set 
according to the theoretical footprint value of this 
pixel relative to the centre of the 100\,$\mu$m 
point-spread function (PSF).

%%%%
%\begin{figure}[h!!!]
%\centerline{\hbox{\epsfig{file=p3wm.eps, width=6.5cm}}}
%\caption[]{Schematic representation of the model 
%apertures constructed to investigate the 100\,$\mu$ 
%measurements of the P3 detector. The 6$\times$6 blue 
%array is the C100 camera pixel grid. Different blue
%tones correspond to different weights (see text). 
%The red circles and the red square represent the P3 
%apertures: 79\arcsec, 99\arcsec, 120\arcsec\, circular, 
%127\arcsec$\times$127\arcsec\, rectangular and 180\arcsec\,
%cicular, from the centre, respectively. }
%\end{figure}

\subsection{Covered observing modes}
A detailed description of ISOPHOT's observing modes 
(AOT = Astronomical Observing Template)
can be found in the ISOPHOT Handbook \citep{Laureijs2003}.
Below we summarize the essentials features for the derivation of 
the sky noise. 
Table~\ref{table:tm} contains a traceability matrix of
applicable configurations per AOT.
%%%%%%
\begin{table*}

\begin{tabular}{cccccc}
\hline
ISOPHOT AOT & submode & detector & aperture & number of  & separation \\ 
            &         &          &          & reference positions &  target--reference  \\ \hline   
PHT\,03       & staring  & P3   & 79\arcsec--180\arcsec & 1      & 1{\farcm}5--10\arcmin \\
	      & chopping & P3   & 79\arcsec--180\arcsec & 1--2   & 90\arcsec--180\arcsec \\ 
PHT\,05       & staring  & P3   & 79\arcsec--180\arcsec & 1      & 1{\farcm}5--10\arcmin \\
PHT\,17/18/19 & staring  & P3   & 79\arcsec--180\arcsec & $\ge$1 & 1{\farcm}5--1{\fdg}5 \\  
PHT\,22	   & staring  & C100 & 138\arcsec$\times$138\arcsec & 1  & 2{\farcm}5--10\arcmin \\
           & staring  & C200 & 184\arcsec$\times$184\arcsec & 1  & 3\arcmin--10\arcmin \\
           & chopping & C100 & 46\arcsec$\times$46\arcsec & 1--2  & 135\arcsec--180\arcsec \\
	   & chopping & C200 & 184\arcsec$\times$184\arcsec & 1  & 180\arcsec \\
        & mini-map (3$\times$3) & C100 & 46\arcsec$\times$46\arcsec & 24  & 46\arcsec--130\arcsec \\
	& mini-map (2$\times$2) & C200 & 92\arcsec$\times$92\arcsec & 8  & 92\arcsec--130\arcsec \\
PHT\,25	   & staring  & C100 & 138\arcsec$\times$138\arcsec & 1  & 2{\farcm}5--10\arcmin \\
           & staring  & C200 & 184\arcsec$\times$184\arcsec & 1  & 3\arcmin--10\arcmin \\
PHT\,32	   & chopping oversampled  & C100 & 46\arcsec$\times$46\arcsec\, or 138\arcsec$\times$138\arcsec 
							& $\ge$1  & 46{\arcsec}--138{\arcsec} \\
	   & chopping oversampled  & C200 & 92\arcsec$\times$92\arcsec\, or 184\arcsec$\times$184\arcsec 
	   						& $\ge$1  & 92{\arcsec}--184{\arcsec} \\
PHT\,37/38/39 & staring  & C200 & 184\arcsec$\times$184\arcsec & $\ge$1  & 3\arcmin--1{\fdg}5 \\	   		   	   	   	
\hline
\end{tabular}
\caption[]{Traceability matrix for the configurations of the investigated C100, C200
and P3 Astronomical Observing Templates}
\label{table:tm}
\end{table*}
%%%%%%

\subsubsection{Staring}
Staring observations were mainly performed as on-off measurements.
The on-off distance was freely selectable, but was in most cases
a few arcminutes. There was the possibility of executing 
a sparse map including several reference positions to any of 
several target positions. Staring observations are compatible
with chopping measurements (rectangular or triangular, depending on the
number of reference positions) in our analysis. 

%%%%
\subsubsection{Chopping}
Chopping observations could be performed with:
\begin{itemize}
   \item one target and one reference position 
   	({\it rectangular chopping}, see Figs.~1a--c). Chopper 
	throws had to be chosen in the range of 
	90\arcsec\,$\le$\,$\theta$\,$\le$\,180\arcsec 
	(see Table~\ref{table:tm}). 
   \item one target and two reference positions
	({\it triangular} and {\it sawtooth} chopping, see Fig.~1a--c).
	Since we do not consider any dependence on the direction of the 
	chopper throw relative to the target, sawtooth and triangular 
	chopping are equivalent in our analysis and are represented by 
	{\it triangular} chopping. Typical chopper 
	throws are listed in Table~\ref{table:tm}. 
\end{itemize}
%%%%%%
\subsubsection{Mapping} 
Confusion noise on maps may be determined in many ways, 
depending on the point-source flux extraction method. This 
can be e.g. similar to some kind of chopping (see above)
using single detector pixels as measuring apertures. 
A typical and effective way is the {\it aperture photometry}, 
using a single detector pixel or apertures as target aperture 
and an annulus of pixels 
placed at a specific distance as reference aperture (see Fig.~1d). 

%%%%%%
\subsubsection{Mini-maps}
Mini-maps represented a special observing mode which was basically used for 
observing point sources with high on-source and background redundancy. 
In mini-map mode the detector (C100 or C200
cameras) moved 'around' the source in a way that the source was centered
in each detector pixel once during the measurement (see Figs.~1e and f 
for a schematic representation). The confusion noise was calculated 
according to the steps of the detector motion around a (hypothetic) 
source considering the detector pixel with the source centered in as 
target and the other pixels as reference apertures. All pixels were
assumed to have equal sensitivities.   

%%%%%%
\subsubsection{Oversampled maps (P32)}
Oversampled maps were performed on a regular grid, composed of a series of 
overlapping parallel scans in the spacecraft y-axis direction. 
The chopper was used for oversampling between individual spacecraft 
positions along the scan line. The same celestial position was observed 
during several raster pointings allowing for elimination of temporal changes 
in detector response. Oversamping factors down to 1/3 of the 
C100 and C200 detector pixel size (15\arcsec\, and 30\arcsec, 
respectively) were allowed, therefore the internal pixel size
of oversampled maps is smaller than that of the maps in our database. 
However, the final photometry in an oversampled map can be performed
in a similar way as in the case of P22 staring raster maps 
(see Sect.~2.2.3).  
In this case an aperture size 
corresponding to the ones in our sample (Tables~2--5) should be chosen 
(e.g. C100 pixel or full detector array apertures), i.e. adding up flux values of
the small map pixels. 
%Due to the differences in P22 and P32 maps, these confusion noise 
%estimates for P32 maps are less reliable than in the case of P22 maps.  

%%%%%%%%%%%%%%%%%%%%%%%%%%%%%%%%%%%%%%%%%%%%%%%%%%%%%%%%%%%%%%%%%%%%%%%%%%%%%%%%%%%%%%%%%%%%%

\section{Confusion noise analysis}

\subsection{ISOPHOT maps}
Our database was built from the final maps produced for Paper~I. 
Therefore we refer to this paper for a detailed description of the data 
reduction. In summary, the data reduction comprised the following main steps:
%%%%
\begin{itemize}
\item[--] basic data analysis with PIA\footnote{PIA is a joint
development by ESA Astrophysics Division and the ISOPHOT consortium led 
by the Max-Planck-Institut f\"ur Astronomie (MPIA), Heidelberg}\,V9.0
\citep{Gabriel97} from raw data (integration ramps, ERD) to 
surface brightness calibrated maps (AAP);
\item[--] flat-fielding using first-quartile normalisation;
\item[--] subtraction of the Zodiacal Light emission;
\item[--] calculation of the instrument noise.
\end{itemize} 
%%%%

\subsection{Derivation of the confusion noise}
\begin{figure*}
\hbox{ \epsfig{file=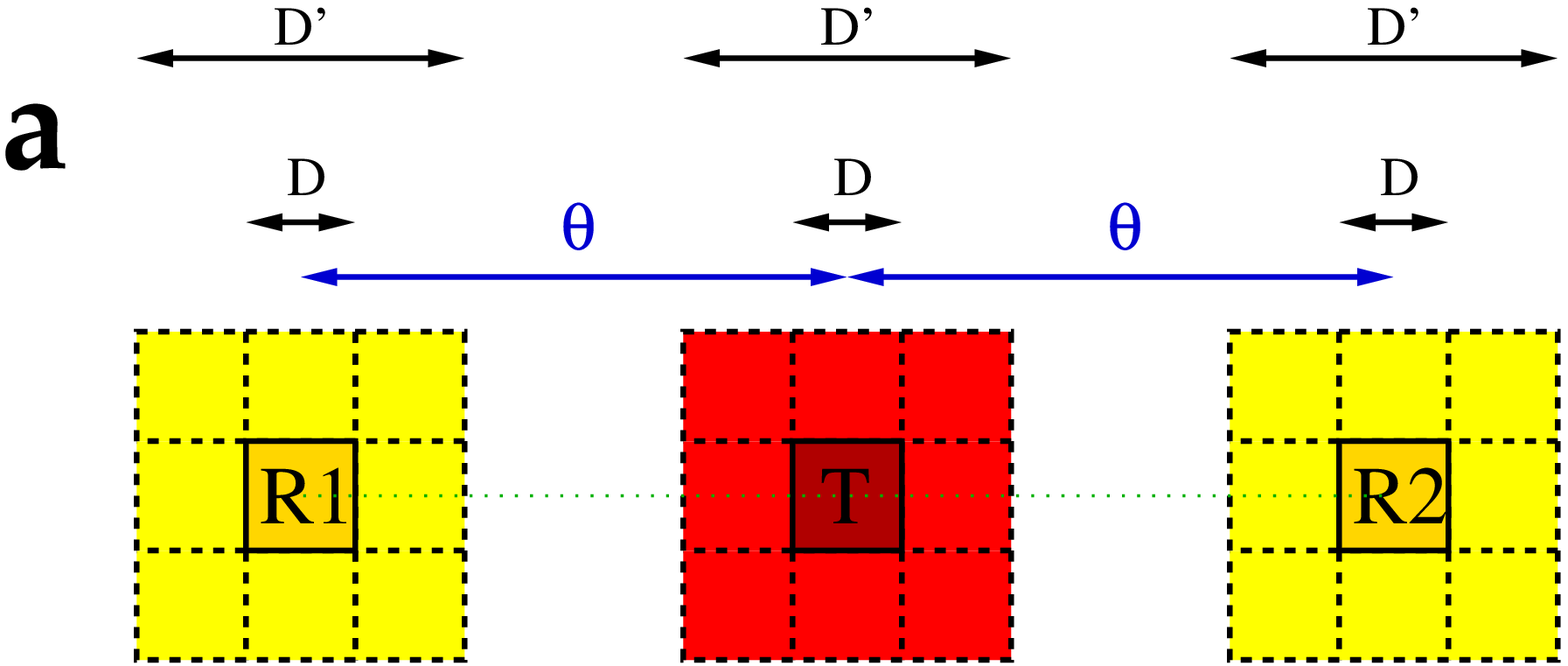, width=7cm}
   \hskip 1.5cm \epsfig{file=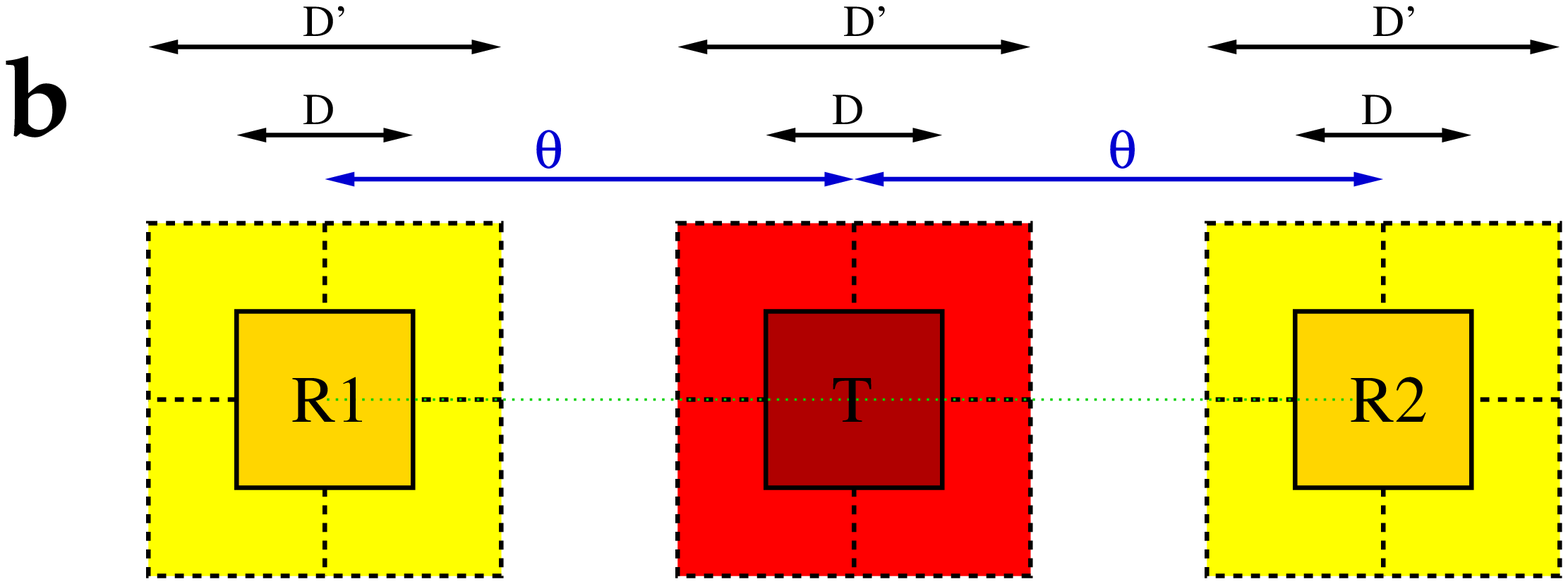, width=8.5cm}}
\vskip 0.6cm   
\hbox{ \epsfig{file=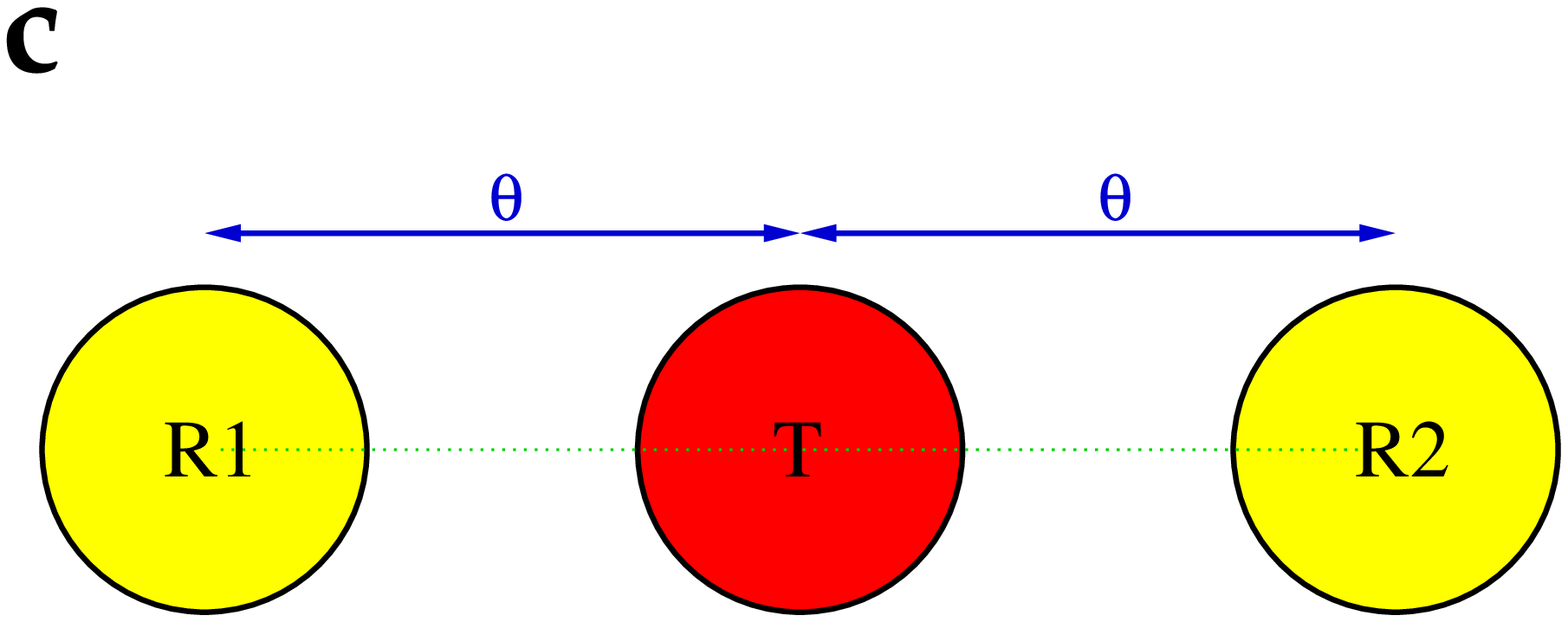, width=7cm}
    \hskip 3cm \epsfig{file=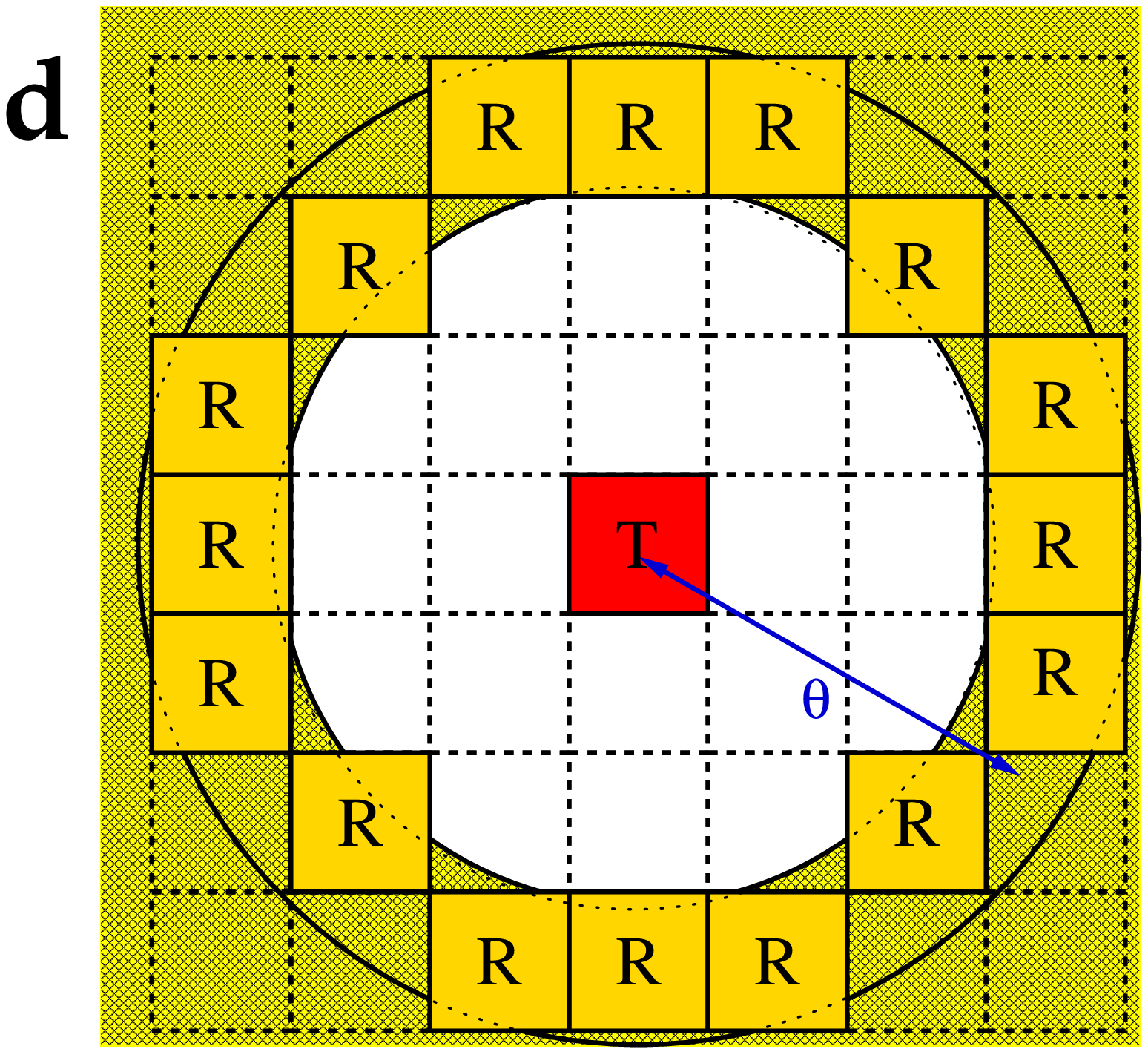, width=6cm}}
\vskip 0.6cm   
\hbox{ \hskip 1.5cm \epsfig{file=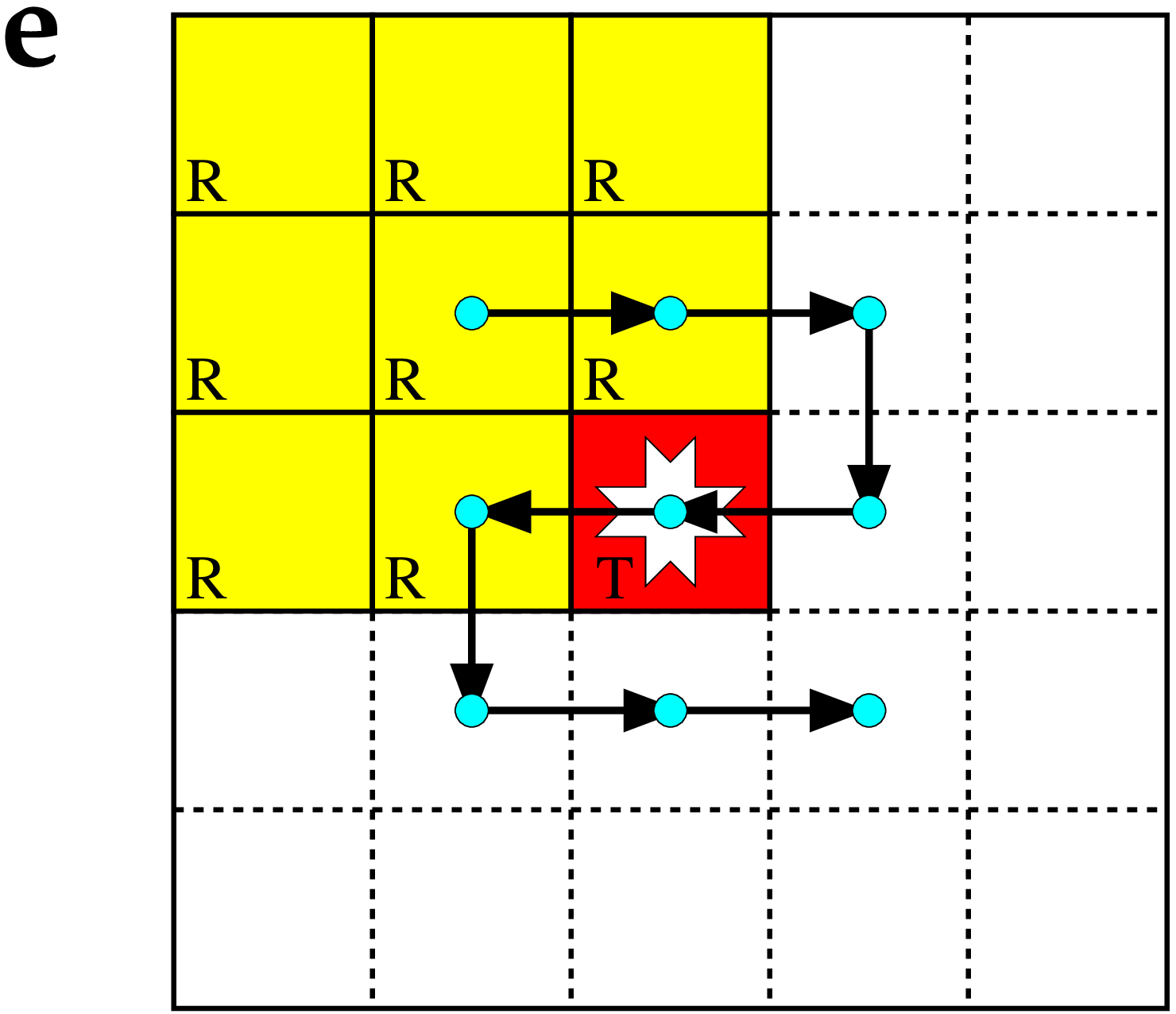, width=5cm}
   \hskip 4cm \epsfig{file=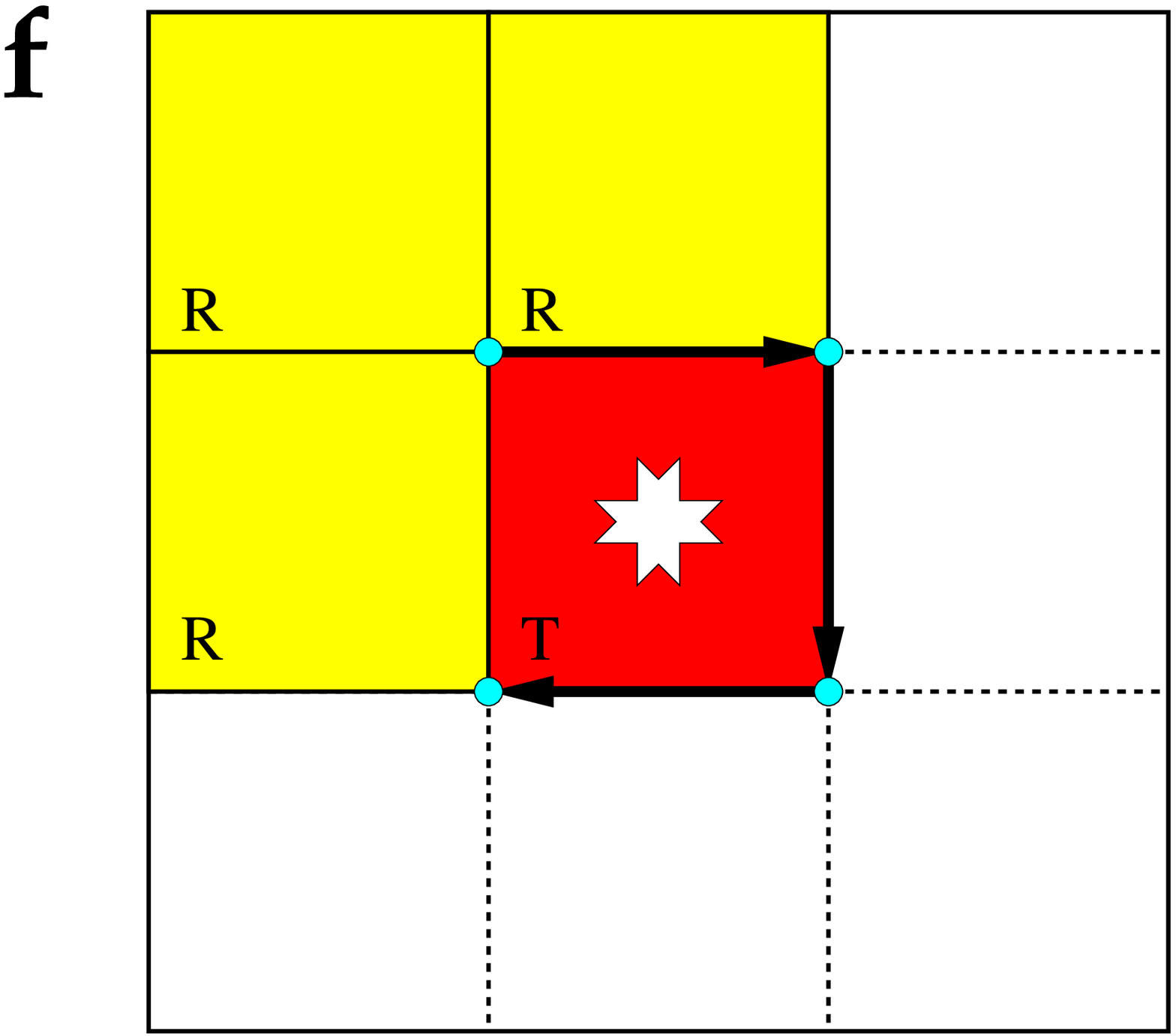, width=5cm}}
%%%%%%        
\caption[]{Measurement configurations investigated in our analysis. 
Target and reference apertures are represented by "T" and "R" flags, 
respectively. In our analysis an aperture can be a single detector pixel 
(P3, C100, C200) or a full detector-array (C100 or C200) as well.
({\bf a}) C100 chopping configurations. For rectangular chopping 
only one reference aperture (R1) and the target aperture (T) is used.
For triangular chopping both R1 and R2 reference apertures are applied. 
Both the target and reference apertures can be single detector pixels
($D$) or full detector arrays ($D'$).  
({\bf b}) The same as (a) but for the C200 detector;
({\bf c}) The same as (a) but for the P3 detector. 
The P3 detector was mainly used in combination with 
circular apertures.   
({\bf d}) Circular (annular) aperture at $\theta$ distance 
around the target aperture. The width of the annulus equals 
the diameter of the target aperture. In practice the annular 
reference aperture is realized by a combination of map detector
pixels as shown in the figure;
({\bf e}) Schematic view of the C100 detector mini-map mode (mini-map 
with 3$\times$3 raster steps). Raster pointings are indicated by the
blue dots. The target and reference apertures for the first raster 
pointing are highlighted in colour; 
({\bf f}) Schematic view of the C200 detector mini-map mode (mini-map with
2$\times$2 raster steps, see C100 description) 
}
\label{fig:config}
\end{figure*}

The confusion noise of a far-infrared map is characterized by the 
{\sl structure function of k$^{th}$}order
\citep[see e.g.][for an introduction]{Gautier,Herbstmeier}:
\begin{equation}
  S(\underline{\theta} ,k) = \Bigg\langle \Bigg| B(\underline{  x}) - 
  {1\over{k}}{\cdot}\sum_k 
   B({\underline{{  x}}}+{\underline{\theta}_k}) 
  {\Bigg|}^2 {\Bigg\rangle}_{\underline{{  x}}}
\end{equation}
where $B$ is the measured sky brightness, $\underline{{  x}}$ is the 
location of the target, $k$ is the number of reference apertures, 
$\underline{\theta}_k$ is the separation
vector of the target and the $k^{th}$ reference aperture and the average
is taken in spatial coordinates over the whole map. 
{$\underline{\theta}_k$}-s are determined by the actual measurement
configuration. Measurement configurations
investigated in this paper are illustrated in Fig.~\ref{fig:config}.
%%%
%%%%%%%%%%%

The structure noise due to the fluctuations of the sky brightness
and instrument noise, $N_{str}$, is defined as:
%%%%%%%
\begin{equation}
N_{str}(\theta,k) = \sqrt{S(\theta,k)}\times\Omega
\end{equation}
%%%%%%%
where $\Omega$ is the solid angle of the measuring aperture.
As shown in Paper~I, the relationship between the 
\emph{confusion noise} and the instrument noise is:
%%%%
\begin{equation}
N_{str}^2 = N^2 + 2N_{inst}^2
\end{equation}
%%%%
where $N$ is the real sky confusion noise and $N_{inst}$ is the 
instrument noise characteristic of a specific far-infrared map. 
We calculated $N$ using the steps (Eqs.~1--3) described above for specific
measurement configurations. 

The final confusion noise values were correlated with the
average surface brightness of the fields, and this relationship
was fitted by a 3-parameter equation (see Paper~I):
\footnotesize
%%%%%
\begin{equation}
  \rm
{{N(\underline{\theta},k,\lambda)}\over{1\,mJy}} 
  = 
  C_0(\underline{\theta},k,{\lambda}) + 
  C_1(\underline{\theta},k,\lambda){\cdot}
  {\bigg\langle {{B}\over{1\,MJy\,sr^{-1}}} \bigg\rangle}^{\eta(\underline{\theta},k,{\lambda})}  
\label{eq:ncoeffs}  
\end{equation}
%%%%%
\normalsize

The $C_0$, $C_1$ and $\eta$ parameters are all functions of
the number ($k$) and configuration ($\underline{\theta}$) of
reference apertures and the wavelength $\lambda$ of the
observation. {  Note that this confusion noise is the superposition 
of extragalactic background and cirrus confusion noise.} 
%%%%%%%

\subsection{Parameter fitting}
The C$_0$, C$_1$ and $\eta$ coefficients were determined
by using a routine based on standard
IDL\footnote{Interactive Data Language, Version 5.4--6.0, Research Systems Inc.}
functions. The routine used the Levenberg-Marquardt 
technique \citep{NumRec} to solve the least-squares problem. 

To perform a completely successful fit for a specific filter/configuration
combination, it is necessary to have data points in the whole surface brightness
range. 
%While C100 90\,$\mu$ and C200 170\,$\mu$m measurements fulfil this
%requirement, this is not the case for the C100 100\,$\mu$ 
%(and therefore P3 100\,$\mu$m, see Sect.~2.4) and C200 200\,$\mu$m
%filters, which lack faint fields. For the latter filters it is impossible
%to derive a proper C$_0$ value from the measured confusion noise values only.
%In these cases the C100 100\,$\mu$m and C200 200\,$\mu$m C$_0$ values
%were calculated from the C100 90\,$\mu$m and C200 170\,$\mu$m C$_0$ values,
%respectively, as described in Sect.~2.6.
Due to the lack of faint fields observed in the C100 100\,$\mu$m and 
C200 200\,$\mu$m filters the C$_0$ parameters for these filters
cannot be fitted properly. Therefore conversions of C$_0$ coefficients
have been applied from C100 90\,$\mu$m to 100\,$\mu$m and 
from C200 170\,$\mu$m to 200\,$\mu$m, as described in Sect.~3.4.

C$_0$ coefficients for the various P3 apertures were 
determined by simulated confusion noise measurements
on synthetic maps with Poissonian brightness-distribution.
In these simulated maps the brightness-scaling was arbitrary,
and only the confusion noise ratio of a single (C100) pixel to a 
simulated P3 aperture was calculated for a specific
configuration. Then these ratios were applied to the 
C$_0$ values of the appropriate configurations in Table~\ref{table:C1_100}
to obtain the C$_0$ coefficients for P3 apertures.

\subsection{Coefficients for non-investigated filters}
For some other filters, which are similary strongly affected by
confusion noise (C100: 105\,$\mu$m, C200: 120, 150 and 180\,$\mu$m) 
the number of available maps was too small to perform the 
investigation as above. 
For these filters we use transformed coefficients 
C$_0$, C$_1$ and $\eta$ of the filters investigated. 
The transformations are done as follows:
\begin{itemize}
\item The C$_0$ and C$_1$ coefficients 
	of the confusion noise -- surface brightness relations
	reflect the spatial structure of the emission. This is
	expressed via the spectral index $\alpha$ 
	\citep{Helou,Gautier} introducing a scaling of 
	$(\lambda_{1}/\lambda_{0})^{1-{{1}\over{2}}\alpha}$. 
\item Since $\alpha\,\approx\,0$ for the extragalactic background, C$_0$
	scales as: C$_0$($\lambda_1$,k,$\theta$)\,=
		\,C$_0$($\lambda_0$,k,$\theta$)$\times(\lambda_{1}/\lambda_{0})$.  
\item {  In Paper II we derived an average cirrus spectral index of
	$\langle\alpha\rangle$\,$\approx$\,--3 for all filters investigated there. 
	We assume here that this can be applied to other wavelengths as well. 
     % C$_1$ scales as $(\lambda_{1}/\lambda_{0})^{1-{{1}\over{2}}\alpha}$. The actual 
     %	value of $\alpha$ for the cirrus emission 
     %	depends on the wavelength and on the surface brightness. However, 
     %	this dependence is not explicitly known for the non-investigated
     %	filters. Therefore, here we use the \emph{canonical} value of 
     % 	$\alpha$ for the cirrus emission, $\alpha_{cirrus}$\,=\,--3.
     %	Using this value, C$_1$ coefficients scale as:
     This leads to the scaling:}
     	C$_1$($\lambda_1$,k,$\theta$)\,=\,C$_1$($\lambda_0$,k,$\theta$)
	  	$\times(\lambda_{1}/\lambda_{0})^{2.5}$.  
\item  
      %  The exponent $\eta$ does not depend explicitly on the spatial structure 
      %  ($\alpha$). $\eta$ is derived from the $B$ -- $P_0$ relation, 
      %  where $B$ is the average surface 
      %  brightness of the field and $P_0$ is the fluctuation power at the 
      %	 reference spatial frequency \citep{Gautier}. Since this relation is  
      %  not known for the non-investigated filters either, we assume, that it has no 
      % wavelength-dependence,
      {  $\eta$ values show a relatively small scatter, 
      and are independent of the filter or even the detector used. 
      The average values for 
      the 90-100\,$\mu$m and for the 170-200\,$\mu$m filters 
      (see Tables~2...6) are $\eta_{100}$\,=1.44$\pm$0.18 and 
      $\eta_{200}$\,=1.58$\pm$0.19, respectively. 
      This implies that either 
      the average values or those of individual configurations can be 
      adopted for predictions 
      at other wavelengths. Since the $\eta$ values of specific 
      configurations may still reflect some individual properties, we
      apply those directly in the transformations,   }
      i.e. $\eta(\lambda_1$,k,$\theta)$\,=\,$\eta(\lambda_0$,k,$\theta)$. 
\end{itemize} 

%%%%%%%%%%
\begin{figure}
\centerline{\epsfig{file=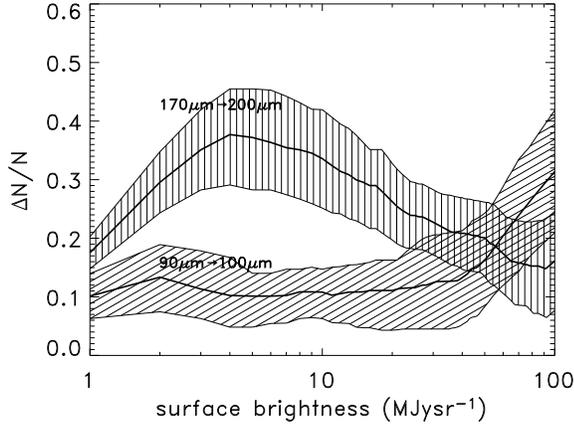, width=8.5cm}}
\caption[]{  Demonstration of the accuracy of confusion noise 
predictions using transformations of the C$_0$, C$_1$ and $\eta$
parameters. The relative uncertainty 
\mbox{$\Delta{N}/N = |N_{tr}-N_{meas}|/N_{meas}$} is plotted versus
surface brightness. $N_{tr}$ is the confusion noise calculated using 
transformed C$_0$, C$_1$ and $\eta$ parameters and $N_{meas}$ is 
the confusion noise calculated using parameters fitted to measurement data
of this specific wavelength. 
The displayed cases are the 90 to 100\,$\mu$m 
and 170 to 200$\mu$m transformations, respectively. The lower and upper
limits of the shaded areas reflect the variations among different 
measurement configurations. }
\label{fig:paramtrans}
\end{figure}
%%%%%%%%%%

{  The usability of these transformed parameters was verified
for those pairs of filters for which a sufficient number of measurements 
was available. For the C100 100\,$\mu$m filter a transformation was 
done from the C100 90\,$\mu$m filter and compared with the 100\,$\mu$m
measurements proper, the same was done for the C200 170\,$\mu$m and 
200\,$\mu$m filter pair. This test was only possible for the C$_1$ 
and $\eta$ parameters. Both the C100 100\,$\mu$m and C200 200\,$\mu$m 
data bases lack low surface brightness measurements and C$_0$ had to be derived 
from the 90\,$\mu$m and 170$\mu$m value, respectively.   
The relative uncertainty 
\mbox{$\Delta{N}/N\,=\,|N_{tr}-N_{meas}|/N_{meas}$}, was determined,
where $N_{tr}$ is the confusion noise using transformed parameters and 
$N_{meas}$ is the confusion noise using parameters fitted to measurement data. 
The results for the 
90-to-100\,$\mu$m and the 170-to-200\,$\mu$m transformations are presented
in Fig.~\ref{fig:paramtrans}. 
%%%
%The uncertainty of the 90-to-100\,$\mu$m 
%transformations is $\sim$10\% in the low-to-medium surface brightness range 
%and it increases to 20--30\% in the high surface brightness range. 
%The uncertainty of the 170-to-200\,$\mu$m transformation is
%typically 20--30\%, has a peak of 40\% at medium surface brightness, 
%but stays below 50\% even in the extreme cases. 
%%%
The fitting of the C$_0$, C$_1$ and 
$\eta$ parameters on the one hand and systematic effects like differences 
in filter bandwidth, changes in the steepness or flatness of the 
sky background spectral energy distribution  
between two filters, etc. on the other hand, introduce uncertainties.
Considering these, our tests proved that the conversion scheme can be effectively 
used to estimate confusion noise for non-investigated filters.
Taking into account the dependence of confusion noise 
on surface brightness and the uncertainty of measuring the latter,
even an uncertainty of 50\% provides 
an acceptable range for the confusion noise estimates. 
%It should be noted that 
%the low surface brightness domain in Fig.~\ref{fig:paramtrans} is highly
%affected by the forced 90-to-100\,$\mu$m and 170-to-200\,$\mu$m 
%C$_0$ values.
}

%%%%%%%%%%%%%%%%%%%%%%%%%%%%%%%%%%%%%%%%%%%%%%%%%%%%%%%%%%%%%%%%%%%%%%%%%%%%%%%%%%%
\section{Results for ISOPHOT filters}
\subsection{Tables and data products}

%------ C1_90 ----------------------------
%%%%%%%%%%%%%%%%%%%%%%%%%%%%%%%%%%%%%%%%%%%%%%%%%%%%%%%%%%%%%%%%%%%%%%%%%%%%
%%%%%%%%%%%%%%%%%%%%%%%%%%%% C1_90 results %%%%%%%%%%%%%%%%%%%%%%%%%%%%%%%%%%
%%%%%%%%%%%%%%%%%%%%%%%%%%%%%%%%%%%%%%%%%%%%%%%%%%%%%%%%%%%%%%%%%%%%%%%%%%%%
\begin{table}[h!!!]
%
%\small-
\begin{tabular}{|ccc|rrr|}
%\scriptsize
%%%%%%%%%%%%%%%%%%%%%%%%%%%%%%%% 90um %%%%%%%%%%%%%%%%%%%%%%%%%%%%%%%%%%%%%%
\hline
  ap. & con. & $\theta$ & C$_0$ & C$_1$ & $\eta$ \\ 
      &      &           & (mJy) & (mJy) &         \\ \hline
       
  P & R & 92\arcsec  & 10.5$\pm$~2.3 &  0.95$\pm$0.11 & 1.47$\pm$0.31 \\ 
  P & R & 138\arcsec &  8.7$\pm$~2.5 &  1.97$\pm$0.26 & 1.27$\pm$0.26 \\
  P & R & 184\arcsec &  8.6$\pm$~2.7 &  2.04$\pm$0.26 & 1.34$\pm$0.26 \\ 
  P & R & 230\arcsec &  8.3$\pm$~2.8 &  2.10$\pm$0.28 & 1.38$\pm$0.26 \\ 
    \hline
  P & T & 92\arcsec  &  9.4$\pm$~1.9 &  0.63$\pm$0.10 & 1.47$\pm$0.33 \\
  P & T & 138\arcsec &  7.3$\pm$~2.0 &  2.09$\pm$0.30 & 1.07$\pm$0.24 \\
  P & T & 184\arcsec &  7.9$\pm$~2.0 &  1.67$\pm$0.19 & 1.25$\pm$0.26 \\
  P & T & 230\arcsec &  6.5$\pm$~2.0 &  2.43$\pm$0.42 & 1.17$\pm$0.24 \\
     \hline
  P & C & 92\arcsec  &  8.1$\pm$~1.8 &  0.74$\pm$0.14 & 1.36$\pm$0.30 \\
  P & C & 138\arcsec &  6.3$\pm$~1.8 &  1.63$\pm$0.28 & 1.11$\pm$0.25 \\
  P & C & 184\arcsec &  6.0$\pm$~1.7 &  1.56$\pm$0.15 & 1.23$\pm$0.26 \\
    \hline
  F & R & 92\arcsec  &  25.4$\pm$14.7 &  9.30$\pm$2.52 & 1.32$\pm$0.18 \\
  F & R & 138\arcsec &  28.4$\pm$19.3 & 13.27$\pm$3.79 & 1.33$\pm$0.17 \\
  F & R & 184\arcsec &  23.8$\pm$21.4 & 17.35$\pm$5.10 & 1.32$\pm$0.16 \\
  F & R & 230\arcsec &  17.9$\pm$21.8 & 20.70$\pm$6.16 & 1.32$\pm$0.15 \\
    \hline
  F & T & 92\arcsec  &  28.6$\pm$10.2 &  2.58$\pm$0.43 & 1.33$\pm$0.26 \\
  F & T & 138\arcsec &  32.6$\pm$15.7 &  6.46$\pm$1.42 & 1.31$\pm$0.18 \\
  F & T & 184\arcsec &  30.6$\pm$16.3 &  8.89$\pm$3.88 & 1.35$\pm$0.35 \\
  F & T & 230\arcsec &  25.5$\pm$18.4 & 12.59$\pm$6.11 & 1.30$\pm$0.34 \\ 
    \hline
%%%%%
\end{tabular}

\normalsize

\caption[]{    
Fitted coefficients of Eq.~\ref{eq:ncoeffs}
for the C100 camera 90$\mu$m filter. Abbreviations in the table:
"ap."\,=\,aperture: P\,=\, single pixel, F\,=\,full array; 
"con."\,=\,configuration for sky reference determinaion:
R\,=\,rectangular chopping (one reference position), 
T\,=\,triangular chopping (two reference positions), 
C\,=\,annular aperture. 
%$^{*}$ Zero C$_0$ values of F~T~184\arcsec and 
%230\arcsec configurations are caused by the negligible contribution of
%the surface brightness inpenendent confuision noise component
%(C$_0$) relative to the cirrus part 
%(C$_1 \times \langle$B$\rangle ^{\eta}$), as described in Sect.~2.6.1.
%
}
\label{table:C1_90}
\end{table}

%%%%%%%

%------ C1_100 ---------------------------
%%%%%%%%%%%%%%%%%%%%%%%%%%%%%%%%%%%%%%%%%%%%%%%%%%%%%%%%%%%%%%%%%%%%%%%%%%%%
%%%%%%%%%%%%%%%%%%%%%%%%%%%% C1_100 results %%%%%%%%%%%%%%%%%%%%%%%%%%%%%%%%%%
%%%%%%%%%%%%%%%%%%%%%%%%%%%%%%%%%%%%%%%%%%%%%%%%%%%%%%%%%%%%%%%%%%%%%%%%%%%%
\begin{table}[h!!!]
%
%\small-
\begin{tabular}{|ccc|rrr|}
%%%%%%%%%%%%%%%%%%%%%%%%%%%%%% 100 um %%%%%%%%%%%%%%%%%%%%%%%%%%%%%%%%%%%%%%%
\hline
  ap. & con. & $\theta$ & C$_0$ & C$_1$ & $\eta$ \\ 
      &      &           & (mJy) & (mJy) &         \\ \hline
  P & R & 92\arcsec  & 11.6$\pm$~2.5 &  1.35$\pm$0.71 & 1.39$\pm$0.11 \\
  P & R & 138\arcsec &  9.6$\pm$~2.8 &  1.47$\pm$0.67 & 1.44$\pm$0.12 \\
  P & R & 184\arcsec &  9.5$\pm$~3.0 &  1.52$\pm$0.72 & 1.51$\pm$0.13 \\
  P & R & 230\arcsec &  9.2$\pm$~3.1 &  1.60$\pm$0.75 & 1.55$\pm$0.15 \\ \hline
  P & T & 92\arcsec  & 10.3$\pm$~2.1 &  1.14$\pm$0.69 & 1.29$\pm$0.08 \\
  P & T & 138\arcsec &  8.1$\pm$~2.2 &  1.44$\pm$0.70 & 1.28$\pm$0.12 \\
  P & T & 184\arcsec &  8.7$\pm$~2.3 &  1.42$\pm$0.70 & 1.43$\pm$0.12 \\
  P & T & 230\arcsec &  7.2$\pm$~2.2 &  1.57$\pm$0.75 & 1.58$\pm$0.14 \\ \hline
  P & C & 92\arcsec  &  9.0$\pm$~2.0 &  1.13$\pm$0.60 & 1.25$\pm$0.08 \\
  P & C & 138\arcsec &  7.0$\pm$~2.0 &  1.47$\pm$0.70 & 1.25$\pm$0.12 \\
  P & C & 184\arcsec &  6.6$\pm$~1.9 &  1.47$\pm$0.71 & 1.34$\pm$0.13 \\ \hline
  F & R & 92\arcsec  &  28.6$\pm$16.3 &  6.47$\pm$2.03 & 1.51$\pm$0.40 \\
  F & R & 138\arcsec &  31.6$\pm$21.4 &  8.26$\pm$2.53 & 1.55$\pm$0.41 \\
  F & R & 184\arcsec &  25.6$\pm$23.7 &  9.56$\pm$3.34 & 1.59$\pm$0.42 \\
  F & R & 230\arcsec &  19.9$\pm$24.2 & 10.33$\pm$3.67 & 1.61$\pm$0.43 \\ \hline
  F & T & 92\arcsec  &  31.9$\pm$11.3 &  3.42$\pm$0.98 & 1.43$\pm$0.45 \\
  F & T & 138\arcsec &  36.2$\pm$17.5 &  3.71$\pm$0.57 & 1.63$\pm$0.70 \\
  F & T & 184\arcsec &  34.0$\pm$18.1 &  7.38$\pm$2.41 & 1.47$\pm$0.39 \\
  F & T & 230\arcsec &  25.8$\pm$20.4 & 14.03$\pm$4.15 & 1.31$\pm$0.26 \\ \hline
\end{tabular}
\caption[]{     
Fitted coefficients of Eq.~\ref{eq:ncoeffs}
for the 100$\mu$m filter of the C100 detector. 
C$_0$ coefficients have been transformed from 90\,$\mu$m results
(see Sect.~2.5). For the meaning of labels for apertures (ap.)
and configurations (con.) see the caption of Table~2.}
\label{table:C1_100}
\end{table}
%%%%%%%

%------ C2_170 ---------------------------
%%%%%%%%%%%%%%%%%%%%%%%%%%%%%%%%%%%%%%%%%%%%%%%%%%%%%%%%%%%%%%%%%%%%%%%%%%%%
%%%%%%%%%%%%%%%%%%%%%%%%%%%% C100 results %%%%%%%%%%%%%%%%%%%%%%%%%%%%%%%%%%
%%%%%%%%%%%%%%%%%%%%%%%%%%%%%%%%%%%%%%%%%%%%%%%%%%%%%%%%%%%%%%%%%%%%%%%%%%%%
\begin{table}[h!!!]
%
%\small-
\begin{tabular}{|ccc|rrr|}
%\scriptsize
%%%%%%%%%%%%%%%%%%%%%%%%%%%%%%%% 90um %%%%%%%%%%%%%%%%%%%%%%%%%%%%%%%%%%%%%%
\hline
  ap. & con. & $\theta$ & C$_0$ & C$_1$ & $\eta$ \\
  	&	&	   & (mJy)  & (mJy) &         \\ \hline
  P & R & 184\arcsec & 19.8$\pm$~5.4 &  0.94$\pm$0.42 & 1.86$\pm$0.21 \\
  P & R & 276\arcsec & 16.6$\pm$~6.2 &  2.27$\pm$0.54 & 1.71$\pm$0.20 \\
  P & R & 368\arcsec & 12.3$\pm$~6.8 &  3.74$\pm$0.58 & 1.61$\pm$0.19 \\
  P & R & 460\arcsec &  5.6$\pm$~7.5 &  5.87$\pm$0.61 & 1.54$\pm$0.18 \\ \hline 
  P & T & 184\arcsec &  10.2$\pm$~4.5 &  3.01$\pm$0.31 & 1.47$\pm$0.16 \\
  P & T & 276\arcsec &  18.3$\pm$~7.0 &  1.61$\pm$0.22 & 1.66$\pm$0.12 \\
  P & T & 368\arcsec &  18.9$\pm$~6.7 &  1.58$\pm$0.24 & 1.70$\pm$0.12 \\
  P & T & 460\arcsec &  14.4$\pm$~7.6 &  2.06$\pm$0.55 & 1.66$\pm$0.11 \\ \hline
  P & C & 184\arcsec &   9.3$\pm$~6.7 &  3.37$\pm$1.01 & 1.46$\pm$0.17 \\
  P & C & 276\arcsec &  16.4$\pm$~8.5 &  1.83$\pm$0.15 & 1.65$\pm$0.23 \\
  P & C & 368\arcsec &  13.8$\pm$~6.8 &  1.21$\pm$0.50 & 1.72$\pm$0.21 \\ \hline
  F & R & 184\arcsec &  49.1$\pm$21.0 &  4.72$\pm$1.15 & 1.78$\pm$0.21 \\
  F & R & 276\arcsec &  40.4$\pm$24.2 & 10.28$\pm$1.29 & 1.66$\pm$0.20 \\
  F & R & 368\arcsec &  20.7$\pm$28.2 & 17.56$\pm$1.38 & 1.57$\pm$0.19 \\
  F & R & 460\arcsec &   7.5$\pm$27.8 & 21.15$\pm$1.41 & 1.59$\pm$0.19 \\ \hline
  F & T & 184\arcsec &  51.3$\pm$17.6 &  1.40$\pm$0.51 & 1.89$\pm$0.23 \\
  F & T & 276\arcsec &  59.2$\pm$32.4 &  2.01$\pm$0.54 & 1.93$\pm$0.24 \\
  F & T & 368\arcsec &  53.6$\pm$23.9 &  4.55$\pm$1.17 & 1.76$\pm$0.22 \\
  F & T & 460\arcsec &  36.2$\pm$25.4 &  9.19$\pm$1.30 & 1.62$\pm$0.19 \\ \hline
\end{tabular}
\caption[]{      
Fitted coefficients of Eq.~\ref{eq:ncoeffs}
for the 170$\mu$m filter of the C200 camera.
For the meaning of labels for apertures (ap.)
and configurations (con.) see the caption of Table~2.}
\label{table:C2_170}
\end{table}
%%%%

%------ C2_200 ---------------------------
%%%%%%%%%%%%%%%%%%%%%%%%%%%%%%%%%%%%%%%%%%%%%%%%%%%%%%%%%%%%%%%%%%%%%%%%%%%%
%%%%%%%%%%%%%%%%%%%%%%%%%%%% C2_200 results %%%%%%%%%%%%%%%%%%%%%%%%%%%%%%%%%%
%%%%%%%%%%%%%%%%%%%%%%%%%%%%%%%%%%%%%%%%%%%%%%%%%%%%%%%%%%%%%%%%%%%%%%%%%%%%
\begin{table}[h!!!]
%
%\small-
\begin{tabular}{|ccc|rrr|}
%\scriptsize
%%%%%%%%%%%%%%%%%%%%%%%%%%%%%% 200 um %%%%%%%%%%%%%%%%%%%%%%%%%%%%%%%%%%%%%%%
\hline
  ap. & con. & $\theta$ & C$_0$ & C$_1$ & $\eta$ \\ 
        &       &          & (mJy) & (mJy) &          \\ \hline  
  P & R & 184\arcsec &  23.3$\pm$~6.3 &   7.81$\pm$2.35 & 1.42$\pm$0.32 \\
  P & R & 276\arcsec &  19.5$\pm$~7.3 &  10.03$\pm$3.23 & 1.46$\pm$0.32 \\
  P & R & 368\arcsec &  14.5$\pm$~8.1 &  11.05$\pm$3.64 & 1.48$\pm$0.33 \\
  P & R & 460\arcsec &   6.6$\pm$~8.8 &  12.34$\pm$4.16 & 1.49$\pm$0.33 \\ \hline
  P & T &  184\arcsec & 12.0$\pm$~5.2 &   7.17$\pm$2.10 & 1.30$\pm$0.29 \\
  P & T &  276\arcsec & 21.6$\pm$~8.2 &   7.95$\pm$2.40 & 1.40$\pm$0.31 \\
  P & T &  368\arcsec & 22.3$\pm$~7.9 &   8.22$\pm$2.52 & 1.45$\pm$0.32 \\
  P & T &  460\arcsec & 16.9$\pm$~8.9 &   9.75$\pm$3.14 & 1.47$\pm$0.33 \\ \hline
  P & C &  184\arcsec & 11.0$\pm$~7.8 &  7.75$\pm$2.32 & 1.26$\pm$0.28 \\
  P & C &  276\arcsec & 19.3$\pm$10.0 &  8.58$\pm$2.65 & 1.37$\pm$0.30 \\
  P & C &  368\arcsec & 16.3$\pm$~8.0 &  6.26$\pm$1.76 & 1.55$\pm$0.34 \\ \hline
  F & R & 184\arcsec &  57.8$\pm$24.7 &  18.20$\pm$6.62 & 1.58$\pm$0.35 \\
  F & R & 276\arcsec &  47.6$\pm$28.4 &  23.71$\pm$8.98 & 1.61$\pm$0.36 \\
  F & R & 368\arcsec &  24.1$\pm$33.2 &  25.68$\pm$9.84 & 1.64$\pm$0.37 \\
  F & R & 460\arcsec &   8.8$\pm$32.7 &  24.95$\pm$9.52 & 1.67$\pm$0.37 \\ \hline
  F & T & 184\arcsec &  60.4$\pm$20.7 &  14.07$\pm$4.88 & 1.46$\pm$0.32 \\
  F & T & 276\arcsec &  69.9$\pm$38.1 &  17.36$\pm$6.26 & 1.55$\pm$0.34 \\
  F & T & 368\arcsec &  63.1$\pm$28.1 &  18.75$\pm$6.87 & 1.63$\pm$0.36 \\
  F & T & 460\arcsec &  42.6$\pm$29.8 &  14.78$\pm$5.36 & 1.74$\pm$0.41 \\ \hline
\end{tabular}
\caption[]{
Fitted coefficients of Eq.~\ref{eq:ncoeffs}
for the 200\,$\mu$m filter of the C200 camera. 
C$_0$ coefficients have been transformed from the 170\,$\mu$m results
(see Sect.~2.5). For the meaning of labels for apertures (ap.)
and configurations (con.) see the caption of Table~2.}
\label{table:C2_200}
\end{table}
%%%%%%%

%------ mini-maps ------------------------
%%%%%%
\begin{table}
\footnotesize
\begin{tabular}{|cc|rrr|}
\hline
detector & filter & 
  \multicolumn{1}{c}{C$_0$}& 
  \multicolumn{1}{c}{C$_1$}& 
  \multicolumn{1}{c|}{$\eta$} \\ 
  & & \multicolumn{1}{c}{(mJy)} & 
  \multicolumn{1}{c}{(mJy)} & \\ \hline
%  C100 & 90\,$\mu$m &
%      10.8$\pm$3.3 &  1.24$\pm$0.18 & 1.27$\pm$0.09 \\
%  C100 & 100\,$\mu$m &
%               12.0$\pm$3.5 &  2.28$\pm$0.24 & 1.21$\pm$0.10 \\
%%%
  C100 & 90\,$\mu$m &
      8.1$\pm$2.0 &  0.74$\pm$0.18 & 1.36$\pm$0.09 \\
%%%      
  C100 & 100\,$\mu$m &
               9.0$\pm$3.1 &  0.96$\pm$0.24 & 1.36$\pm$0.10 \\ 
%%%	       
  C200 & 170\,$\mu$m &
      12.2$\pm$3.5 &  0.65$\pm$0.12 & 1.63$\pm$0.32 \\
%%%      
  C200 & 200\,$\mu$m &
               14.4$\pm$3.8 & 2.35$\pm$0.98 & 1.41$\pm$0.19 \\
%%%	       
\hline
\end{tabular}
\small
\caption[]{Fitted coefficients of Eq.~\ref{eq:ncoeffs}
for the C100/90\,$\mu$m, C100/100\,$\mu$m, C200/170\,$\mu$m, C200/200\,$\mu$m
detector/filter combinations for the simulated mini-map observing mode.
C$_0$ coefficients for the 100\,$\mu$m filter have been transformed
from the 90\,$\mu$m results, for the 200\,$\mu$m filter from the
170\,$\mu$m results.}
\label{table:minimapres}
\end{table}
%%%%%

%------ P3 -------------------------------
%%%%%%
\begin{table}
\begin{tabular}{|rcc|rrr|}
\hline
%%%%%%%%%%%%%%%%%%%%%%%%%%%%%%%%%%%%%%%%%%%%%%%%%%%%%%%%%%%%%%%%%%%%%%%%%%%%%%%%%%%%%%%%
%%%%%%%%%%%%%%%%%%%%%%%%%%%%%%%%%%%%%%%%%%%%%%%%%%%%%%%%%%%%%%%%%%%%%%%%%%%%%%%%%%%%%%%%
aper. &con.& $\theta$ & C$_0$ & C$_1$ & $\eta$  \\ 
         &     &          & (mJy) & (mJy) &   \\ \hline
%  &  & \multicolumn{3}{c|}{$\theta$=90\arcsec, 105\arcsec} \\ \hline
%%%%%%%%%%%%%%%%%%%%%% P3 model apertures %%%%%%%%%%%%%%%%%%%%%%%%%%  
%%%%%%%%%%%%%%%%%%%%%% theta = 90"/105" %%%%%%%%%%%%%%%%%%%%%%%%%%%%%%
 79\arcsec$\bigcirc$  & R & 90\arcsec & 
 	7.8$\pm$1.7 & 2.25$\pm$0.52 & 1.15$\pm$0.19 \\
  
 79\arcsec$\bigcirc$  & T &  90\arcsec & 
 	6.0$\pm$1.2 & 1.08$\pm$0.19  &  1.13$\pm$0.15  \\

 99\arcsec$\bigcirc$  & T & 90\arcsec &  
 	9.0$\pm$1.9  & 0.74$\pm$0.12 & 1.18$\pm$0.18 \\

\hline
%%%%%%%%%%%%%%%%%%%%%%%%%%%%%%%%%%%%%%%%%%%%%%%%%%%%%%%%%%%%%%%%%%%%%%%%%%%%%%%%%%%%%%%%%%%
%%%%%%%%%%%%%%%%%%%%%%%%%%%%%%%%%%%%%%%%%%%%%%%%%%%%%%%%%%%%%%%%%%%%%%%%%%%%%%%%%%%%%%%%%%%
\hline
%aperture &conf.& $\theta$ & C$_0$ & C$_1$ & $\eta$  \\ \hline
%  & & \multicolumn{3}{c|}{$\theta$=120\arcsec}  \\ \hline
%%%%%%%%%%%%%%%%%%%%%% P3 model apertures %%%%%%%%%%%%%%%%%%%%%%%%%%  
% 79\arcsec$\bigcirc$  & R  120\arcsec & 
%    &\multicolumn{3}{c|}{\colorbox{yellow}{\parbox{32mm}{$~~~~$}}} \\
  
 79\arcsec$\bigcirc$  & T & 120\arcsec &  
    6.0$\pm$1.2 & 1.93$\pm$0.37 & 1.10$\pm$0.18 \\
 
 99\arcsec$\bigcirc$  & R &  120\arcsec &      
    10.9$\pm$2.3 & 2.86$\pm$0.27 & 1.16$\pm$0.19 \\
 
 99\arcsec$\bigcirc$  & T & 120\arcsec &  
    9.0$\pm$1.9 & 1.59$\pm$0.22 & 1.13$\pm$0.19 \\
 
 120\arcsec$\bigcirc$ & R & 120\arcsec & 
    14.0$\pm$3.0 & 2.46$\pm$0.17 & 1.17$\pm$0.18 \\
  
 120\arcsec$\bigcirc$ & T & 120\arcsec &  
    11.7$\pm$2.4 & 1.29$\pm$0.11 & 1.15$\pm$0.17 \\ 

\hline
%%%%%%%%%%%%%%%%%%%%%%%%%%%%%%%%%%%%%%%%%%%%%%%%%%%%%%%%%%%%%%%%%%%%%%%%%%%%%%%%%%%%%%%%%%%
%%%%%%%%%%%%%%%%%%%%%%%%%%%%%%%%%%%%%%%%%%%%%%%%%%%%%%%%%%%%%%%%%%%%%%%%%%%%%%%%%%%%%%%%%%%
\hline
%aperture &conf.& $\theta$ & $\theta$
%   C$_0$ & C$_1$ & $\eta$  \\ \hline
%   &  & \multicolumn{3}{c|}{$\theta$=180\arcsec}  \\ \hline
%%%%%%%%%%%%%%%%%%%%%% P3 model apertures %%%%%%%%%%%%%%%%%%%%%%%%%%  
 79\arcsec$\bigcirc$  & R & 180\arcsec & 
    7.8$\pm$1.7 & 3.30$\pm$0.37  & 1.21$\pm$0.20\\     
 99\arcsec$\bigcirc$  & R & 180\arcsec & 
    10.9$\pm$2.3 & 3.37$\pm$0.34 & 1.18$\pm$0.19 \\
 120\arcsec$\bigcirc$ & R & 180\arcsec & 
    14.0$\pm$3.0 &  3.13$\pm$0.26 & 1.17$\pm$0.18 \\
 127\arcsec$\square$  & R & 180\arcsec & 
     17.1$\pm$4.2 & 3.26$\pm$0.51 & 1.23$\pm$0.21 \\
 180\arcsec$\bigcirc$ & R & 180\arcsec & 
    24.0$\pm$5.2 & 2.33$\pm$0.30 & 1.23$\pm$0.20 \\
\hline
%%%%%%%%%%%%%%%%%%%%%%%%%%%%%%%%%%%%%%%%%%%%%%%%%%%%%%%%%%%%%%%%%%%%%%%%%%%%%%%%%%%%%%%%%%%
%%%%%%%%%%%%%%%%%%%%%%%%%%%%%%%%%%%%%%%%%%%%%%%%%%%%%%%%%%%%%%%%%%%%%%%%%%%%%%%%%%%%%%%%%%%
\hline
%aperture &conf.& $\theta$ & C$_0$ & C$_1$ & $\eta$  \\ \hline
%   &  & 
%  \multicolumn{3}{c|}{$\theta$=240\arcsec} \\ \hline
%%%%%%%%%%%%%%%%%%%%%% P3 model apertures %%%%%%%%%%%%%%%%%%%%%%%%%%  
 120\arcsec$\bigcirc$ & R & 240\arcsec & 
    14.0$\pm$3.0 & 3.61$\pm$0.34 & 1.18$\pm$0.20 \\ 
\hline
%%%%%%%%%%%%%%%%%%%%%%%%%%%%%%%%%%%%%%%%%%%%%%%%%%%%%%%%%%%%%%%%%%%%%%%%%%%%%%%%%%%%%%%%%%%
%%%%%%%%%%%%%%%%%%%%%%%%%%%%%%%%%%%%%%%%%%%%%%%%%%%%%%%%%%%%%%%%%%%%%%%%%%%%%%%%%%%%%%%%%%%

%%%%%%%%%%%%%%%%%%%%%%%%%%%%
\end{tabular}
\small
\caption[]{Fitted coefficients of Eq.~\ref{eq:ncoeffs}
for the simulated P3 detector -- aperture combinations in the 
100\,$\mu$m filter.
Coefficients were derived for only those combinations, which were 
selected for observations .}
\label{table:P3}
\end{table}
%%%%%

%%%%

The fitted C$_0$, C$_1$ and $\eta$ parameters 
for a specific detector / filter / configuration 
combination are tabulated in 
Tables~\ref{table:C1_90}--\ref{table:P3}. 
It should be emphasized that the presented confusion noise values are 
{\it 'per beam'} and {\it 1$\sigma$} values.  
They have to be corrected with the appropriate PSF-fraction
of the beam when compared with point-source confusion noise values.
An example of the functional behaviour of the confusion noise 
with surface brightness is given in fig.~2 in Paper~I. 

Due to the surface brightness range of the measurements 
used in our analysis the predictions with the coefficients 
in Tables~2--7 are reliable in the 
1\,$\le$\,$\langle B \rangle$\,$\le$\,100\,MJy\,sr$^{-1}$ 
and in the 
1\,$\le$\,$\langle B \rangle$\,$\le$\,200\,MJy\,sr$^{-1}$ 
range for the C100/P3 and the C200 detectors,
respectively.  There are no suitable measurements
at high surface brightness and in addition 
the structure of the FIR emission may change
significantly above this level 
(see Kiss et al., 2003, hereafter Paper~II).
Therefore we cannot give accurate 
estimates for very high surface brightness 
values using our current database.  

Based on the results compiled in the tables we 
constructed all-sky confusion
noise maps, one for each detector / filter / configuration combination.
Surface brightness values {  for all positions of our 1\degr\,resolution grid} 
were derived from COBE/DIRBE data 
with the Zodiacal Light contribution removed and  
interpolated to the ISOPHOT bands, as described in Paper~I.
The only difference is that the cirrus colour temperature was
not fixed to 20\,K, but was derived from the COBE/DIRBE
100, 140 and 240\,$\mu$m surface brightness values.
%%%%%
{  Due to the in general higher uncertainty a lower weight 
was given to the 140\,$\mu$m band data. The long 
wavelength baseline of the 100 and 240\,$\mu$m bands and
the COBE/DIRBE surface brightness accuracies of a few percent
provide a final temperature accuracy $\delta$T\,$\le$\,0.2\,K.

We calculated the confusion noise from these surface 
brightness values using Eq.~4 
and the C$_0$, C$_1$ and $\eta$ parameters corresponding to the 
actual measurement configuration (Tables~2...7). }
%%%%% 
%A transformation was applied between the COBE/DIRBE and 
%ISOPHOT photometric systems as described in Sect.~5.2.2.
The maps are available in the electronic version of this 
paper\footnote{http://kisag.konkoly.hu} in FITS format. 
The data format is described in Appendix~A.1. 
One advantage of these maps is that they are based on a 
homogeneous all-sky surface brightness calibration. 
In quite a number of cases 
the determination of the background value from the ISOPHOT
measurement itself may be less accurate (e.g. in chopped observations). 
In this case the confusion noise estimated by using 
Eq.~4 in combination with the measured background brightness is
unreliable. Our all-sky maps provide information on the average 
confusion noise around the target field, averaged over an 
area of $\sim$1\,deg$^2$,which can be more reliable, if there are no 
strong gradients in the background brightness over this scale.  
There are also maps for the filters originally not investigated, 
with the transformations described in Sect.~3.4 applied. 

\subsection{Noise performance of various observing modes}

%%%%
\begin{figure}
\centerline{\epsfig{file=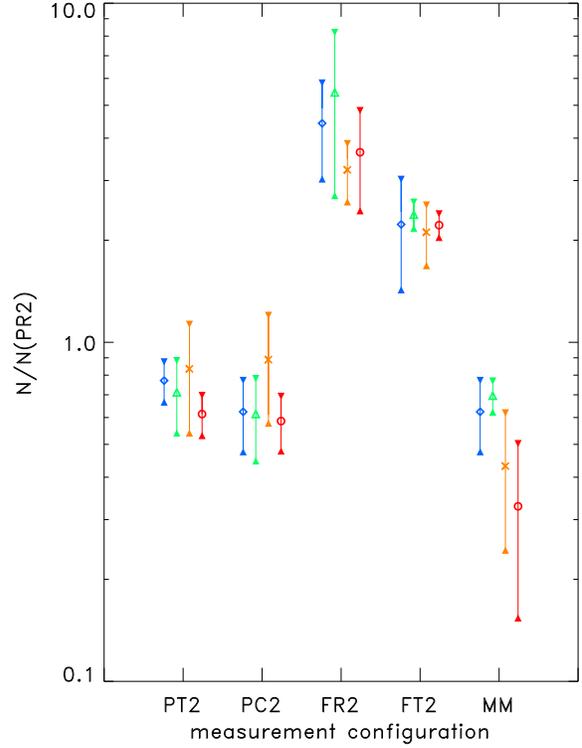, width=8.5cm}}
\caption[]{Comparison of the confusion noise values for different
observing modes and measurement configurations. The measurement 
configurations are coded by the labels and the separation $\theta$ 
is given in detector pixels of the actual detector:
PT2 -- triangular chopping, pixel aperture, $\theta$\,=\,2;
PC2 -- circular aperture (annulus of pixels), $\theta$\,=\,2;
FR2 -- rectangular chopping, detector array aperture, $\theta$\,=\,2;
FT2 -- triangular chopping, detector array aperture, $\theta$\,=\,2;
MM -- mini-map mode. 
All values are given as ratios with regard to the reference configuration
'PR2': rectangular chopping, single pixel aperture, $\theta$\,=\,2.  
The ISOPHOT filters are marked by the following symbols:
diamond: C100 90\,$\mu$m; triangle: C100, 100\,$\mu$m; 
asterisk: C200. 170\,$\mu$m; circle: C200, 200\,$\mu$m;
The vertical bars limited by the arrowheads represent the range of 
ratios for surface brightness values of 
2\,MJy\,sr$^{-1}$\,$\le$\,$\langle$B$\rangle$\,$\le$\,100\,MJy\,sr$^{-1}$.}

\label{fig:allcomp}
\end{figure}
%%%%

In Fig.~\ref{fig:allcomp} we compare the confusion noise values  
obtained for various observing modes. We have chosen the rectangular
chopping with a single pixel aperture and with a separation of 
$\theta_{min}$\,=\,2\,pixels as reference (denoted 'PR2'). 
This is compared with the results of other configurations
(detailed in the figure caption). 
 
The background determination in mini-map mode turned out to provide the 
lowest confusion noise in most cases. 
The second reference position in triangular chopping reduces 
the confusion noise significantly (by 15--50\%) 
compared to rectangular chopping. Full detector array
apertures prove to be 3--5 times more strongly affected by confusion noise
than the same configuration with single pixels. 

\subsection{Noise characterization of individual measurements 
in the ISO Data Archive}

The confusion noise can be a severe limit both for the signal-to-noise
and the photometric accuracy of faint sources. 
Due to the analysis described here 
it is now possible to determine a robust confusion noise estimate
for all FIR ISOPHOT measurements performed with the AOTs listed 
in Table~1 and which are compact source measurements including 
background reference positions. From ISO Data Archive (Kessler et al., 2003)
Version~7 onward the Data Quality Report of these observations 
flags the possible cirrus confusion contamination and an associated 
catalogue file gives the confusion noise numbers calculated 
both for the measured ISOPHOT and the COBE/DIRBE background values.
{  These can be directly compared with source fluxes extracted 
from the data products.} 

%%%%%%%%%%%%%%%%%%%%%%%%%%%%%%%%%%%%%%%%%%%%%%%%%%%%%%%%%%%%%%%%%%%%%%%%%%%%%%%%%%%%%%%%%

%%%%%%%%%%%%%%%%%%%%%%%%%%%%%%%%%%%%%%%%%%%%%%%%%%%%%%%%%%%%%%%%%%%%%%%%%%%%%%%%%%%%%%%%%
%%%%%%%%%%%%%%%%%%%%%% SIMULATED MAPS -- PREDICTIONS FOR SIRTF, ASTRO-F, PACS %%%%%%%%%%%

\section{Cirrus confusion noise predictions for infrared space telescopes}

\subsection{Simulated maps of cirrus structures} 

%%%%
{  In this analysis we concentrate on the cirrus component 
of the confusion noise and its scaling between instruments with
various spatial resolutions. Here we consider neither 
the contribution of the extragalactic background nor
the impact by any instrumental effect.}
%%%%
The main characteristic of the cirrus emission at a specific 
wavelength is its spatial structure. This is usually described by 
the spectral index, $\alpha$, of the power spectrum 
of the image, averaged over annuli (see Paper~II for a summary). 
{  With this parameter the power spectrum is $P(f)=P_0(f/f_0)^{\alpha}$, 
where $P(f)$ is the power at the spatial frequency $f$ and $P_0$ 
is the power at the reference spatial frequency $f_0$.}
As was shown in Paper~II, $\alpha$ may vary with wavelength and 
surface brightness. Previous studies (Papers I \& II) unraveled the
structure of the emission for the scale down to the ISOPHOT resolution.
Extrapolations to better spatial resolutions 
of future space telescopes
can be performed using these results and assuming that 
the general behaviour of the spatial structure remains unchanged for
higher spatial frequencies, i.e. the same fractal dimension /  
the same spectral index is valid. 
%%%

{  The correct simulation of the entire cirrus structure of a 
specific sky area in the Fourier 
space would demand the reproduction of both the power and phase
information. However, in any autocorrelation analysis 
like the confusion noise calculation, only the Fourier power is 
important, since the two-dimensional autocorrelation function 
is related to the two-dimensional power spectrum only:
\begin{equation}
C(\theta) = {{1}\over{2\pi}} \int_{0}^{\infty}
  P(f') \, J_{0}(f'\theta)  \, f' \, df' 
\end{equation}
\noindent where $P(f')$ is the two-dimensional power spectrum, 
$J_{0}(x)$ is the circular Bessel-function of the 0$^{th}$ kind
and $\theta$ is the angular separation. The confusion noise 
$N$ is proportional to C($\theta$) or in more complex configurations 
$N$ is a linear combination of some C($\theta_i$)-s.
Different FIR instruments sample the power spectrum at different spatial frequencies
due to their different resolving power, resulting
in different confusion noise levels on the same (cirrus) structure.  
Only $P_0$ depends on the surface brightness $B$ \citep{Gautier}. 
For a power-law type power spectrum the {\it ratio} of two 
power levels at two spatial frequencies is independent of $P_0$ and 
therefore of $B$.  
If one knows the $P_0$--$B$ relation for one specific
instrument (e.g. ISOPHOT), it is straightforward to derive a $P$--$B$ 
relation for another instrument using $P(f)=P_0(f/f_0)^{\alpha}$. This is 
equivalent with a confusion noise -- surface brightness relation for 
this other instrument. 

However, for a specific instrument the size of the detector pixels and the
measurement configuration play an important role as well. Because of this
complexity the easiest way to compare 
different instruments is the confusion noise 
analysis of simulated maps as these would be obtained by these instruments.
This is done in the same way as for the real ISOPHOT maps in Sect.~3.  

%In our analysis we required only that the spectral index
%$\alpha$ of the power spectrum had to be the prescribed one
%(as was shown in Paper II, $\alpha$ may vary in the range 
%\mbox{--2.0\,$\ge$\,$\alpha$\,$\ge$\,--5.0}).
}
%%%
{  
%Maps simulated in this way -- due to their fractal nature -- 
%have no properties through which they could be connected 
%e.g. to the surface brightness of real maps. 
%In this model the confusion noise ratios at two specific separations
%(or at the two corresponding spatial frequencies) 
%are constant for a specific $\alpha$, 
%since the ratio of their corresponding Fourier powers are 
%constant:
%\begin{equation}
%{{P_1}\over{P_2}}\,=\,
%   {{P_0{\big( {{f_1}\over{f_0}} \big) }^{\alpha}}\over{P_0{\big( {{f_2}\over{f_0}} \big) }^{\alpha}}} \,=\,
%   \bigg( {{f_1}\over{f_2}} \bigg)^{\alpha}
%\end{equation}
%\noindent which is independent of $P_0$. If the spectral index of 
%the cirrus were constant everywhere on the sky (e.g. $\alpha$\,=\,--3)
%the surface brightness dependence would be introduced through 
%this P$_0$ parameter only \citep{Helou,Gautier}. If $\alpha$ depends
%on the surface brightness (as we found in Paper~II),  
%a specific $\alpha$ value can be assinged to a specific surface brightness.
%Different FIR instruments -- mainly due to their different resolving power --
%sample the power spectrum at different spatial frequencies, resulting
%in different confusion noise levels on the same (cirrus) structure. 

For ISOPHOT we derived the relation between 
the cirrus confusion noise and the surface brightness of individual
fields in Sect.~4. From simulated fractal maps the confusion noise 
ratios of ISOPHOT and another instrument can be obtained. 
In this way the cirrus confusion noise of any FIR instrument can be
connected to the surface brightness.} 

To perform the investigation we constructed high-resolution
(4096$\times$4096 pixels, 0\farcs5 pixel size) synthetic images for a range of 
spectral index values (--2.0\,$\ge$\,$\alpha$\,$\ge$\,--5.0). 
The generation of the maps is based on the
{\sl random recursive fractal algorithm} by \citet{Elmegreen}. 

%%%%%%%%%%
\begin{figure}
\centerline{\epsfig{file=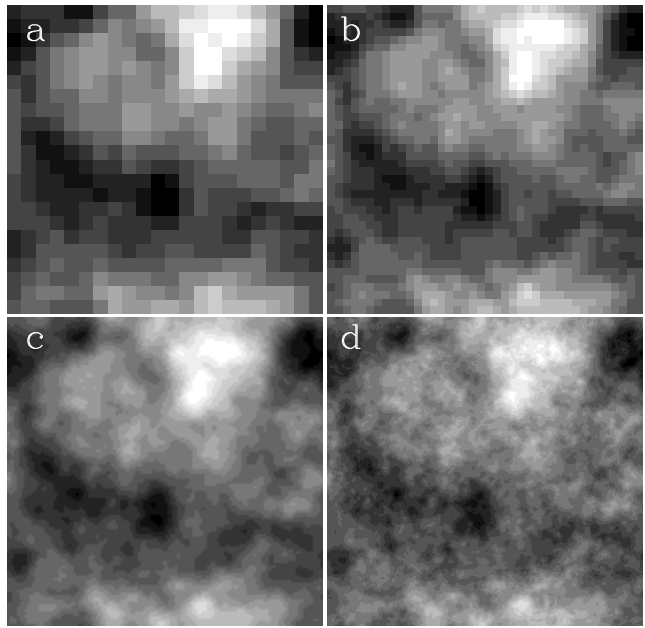, width=8.5cm}}
\caption[]{  Simulated fractal map with $\alpha$\,=\,--3, 
as seen with the spatial resolution of various telescopes/instruments.   
({\bf a}) ISOPHOT 170\,$\mu$m; 
({\bf b}) ASTRO-F/FIS 170\,$\mu$m;
({\bf c}) Spitzer/MIPS 160\,$\mu$m;
({\bf d}) Herschel/PACS 175\,$\mu$m }
\label{fig:simmaps}
\end{figure}
%%%%%%%%%%

The high resolution maps were convolved with the beams 
of the actual telescope/detector/filter combinations and then sampled 
according to the size of the detector pixels. 
The Spitzer/MIPS 160$\mu$m point spread function is available
at the Spitzer Science Center 
website\footnote{http://ssc.spitzer.caltech.edu/mips/psffits/}. 
Point spread functions for the Herschel/PACS 110 and 175\,$\mu$m 
simulated measurements were taken from model calculations
\citep{Okumura+Longval}. 
The ASTRO-F/FIS PSFs were calculated theoretically using the latest 
information on the telescope design \citep{Jeong}.
{  The result of this convolution and resampling is shown in Fig.~\ref{fig:simmaps}.}
   
The confusion noise analysis was performed in the same way 
as for ISOPHOT observations. The analysis was restricted to 
the \emph{triangular chopping} configuration, 
and the separation $\theta$ was chosen to be equal to the 
\emph{resolution limits} of the actual telescope/detector/filter 
combinations. Since confusion noise is especially important 
for detection of faint point sources, we provide 1$\sigma$ point source
confusion noise values after a correction for the footprint  
of the instruments, instead of single-pixel ("per beam") confusion
noise values.

%The surface brightness dependence is introduced via 
%the dependence of the ISOPHOT confusion noise coefficients on the
%surface brightness, as described in Sects.~3.2.1 and 3.2.2

\subsection{Results}
\subsubsection{Conversion coefficients:}
The main parameters of the investigated instruments and 
the conversion coefficients are summarized in Table~\ref{table:phtconv}.
The conversion equation is: 
\begin{equation}
N_{R} = R^{PS} \times N_{PHT}
\label{eq:n2n}
\end{equation}
where $N_{PHT}$ and $N_{R}$ are the confusion noise 
values obtained by ISOPHOT and another instrument 
for the same map, respectively 
and $R^{PS}$ is the conversion coefficient. 
The $R^{PS}$ factor can also be expressed in terms of the 
C$_1$ and central point spread function fraction ($f_{psf}$) 
factors of ISOPHOT and the investigated instrument at their 
resolution limits:
\begin{equation}
R^{PS} = {{C_1^R (\lambda,\theta_{min}^R,k,\alpha){\cdot}f_{psf}^{PHT}}
		\over{C_1^{PHT} (\lambda_{ref},\theta_{min}^{PHT},k,\alpha)}
		{\cdot}f_{psf}^{R}}
\label{eq:R2C1}
\end{equation}
%%%
%\noindent where $C_1^R (\lambda,\theta_{min}^R,k)$ and
%$C_1^{PHT} (\lambda,\theta_{min}^{PHT},k)$
%are the C$_1$ coefficients for the investigated instrument and for ISOPHOT,
%respectively, using a specific configuration. 
%%%%
{  \noindent These R$^{PS}$ factors of Table~8 can be compared with the 
simple telescope resolution scaling ($\lambda/D$)$^{2.5}$ (for $\alpha$\,=\,--3)
in the \citet{Helou} formula. For the 85\,cm Spitzer telescope
R$_{HB}$\,=\,4.2$\cdot$10$^{-1}$ and for the 3.5\,m Herschel telescope  
R$_{HB}$\,=\,1.2$\cdot$10$^{-2}$. An example of a graphical 
comparison is shown in Fig.~5.}

%%%
\begin{table*}
\begin{center}
\begin{tabular}{llccccrr}
\hline\hline
instrument & filter & $R^{PS}({\alpha=-2.0})$ & $R^{PS}({\alpha=-3.0})$ 
  & $R^{PS}({\alpha=-4.0})$ & $R^{PS}({\alpha=-5.0})$ & res. limit & PHT ref. \\ \hline
ASTRO-F/FIS &  170\,$\mu$m & 4.4$\times$10$^{-1}$ & 3.9$\times$10$^{-1}$ 
  & 3.5$\times$10$^{-1}$ & 3.1$\times$10$^{-1}$ & 65{\farcs}5 & 170\,$\mu$m \\
Spitzer/MIPS &  160\,$\mu$m & 1.6$\times$10$^{-1}$ & 1.2$\times$10$^{-1}$ 
  & 8.7$\times$10$^{-2}$ & 6.5$\times$10$^{-2}$ & 46{\farcs}5 & 170\,$\mu$m \\
Herschel/PACS & 110\,$\mu$m & 3.0$\times$10$^{-2}$ & 1.3$\times$10$^{-2}$ 
  & 5.3$\times$10$^{-3}$ & 2.4$\times$10$^{-3}$ & 8{\farcs}5 & 90\,$\mu$m \\ 
Herschel/PACS & 175\,$\mu$m & 1.6$\times$10$^{-2}$ & 7.4$\times$10$^{-3}$ 
  & 3.0$\times$10$^{-3}$ & 1.4$\times$10$^{-3}$ & 13{\farcs}5 & 170\,$\mu$m \\       
\hline\hline
\end{tabular}
\end{center}
\caption[]{Conversion coefficients for \emph{point sources}
between ISOPHOT and the 
specified instruments for cirrus confusion noise calculations. 
The columns of the table contain:
1.) Instrument;
2.) Filter; 
3.) Ratio of confusion noise values at the resolution limits for 
  a spectral index of $\alpha$\,=\,--2.0;
4.) same for $\alpha$\,=\,--3.0;  
5.) same for $\alpha$\,=\,--4.0;  
6.) same for $\alpha$\,=\,--5.0; 
7.) Resolution limits of the current instrument/filter combination
(76\arcsec and 143\arcsec for the ISOPHOT 90 and 170\,$\mu$m filters,
respectively);   
8.) ISOPHOT reference filter.
}
\label{table:phtconv}
\end{table*} 
%%%

\subsubsection{All-sky confusion noise maps:}

Based on the conversion factors derived above we were able to 
produce low spatial resolution all-sky {\it cirrus} 
confusion noise prediction maps {  in a similar way as for 
the ISOPHOT  all-sky confusion noise maps.} 
{  Surface brightness values were derived from COBE/DIRBE data, 
following the same scheme as in Sect.~4.1.  
The confusion noise values of the all-sky maps are calculated
for the configuration PT2 (Tables 2 and 4, 
$\theta$\,=\,92\arcsec~for C100 90\,$\mu$m 
and  $\theta$\,=\,184\arcsec~for C200 170\,$\mu$m )
%The confusion noise values of the all-sky maps are calculated 
%by applying the ISOPHOT {\it cirrus} noise coefficients 
%(PT2, obtained as described in Sect.~4.1) 
%and the conversion factors based on the analysis of 
%the simulated maps (as given in Table~\ref{table:phtconv}) 
%to the monochromatic surface brightness of a specific sky position. 
The C$_0$, C$_1$ and $\eta$ coefficients for this configuration 
cannot be applied directly, since those resulted from the fits
to the total surface brightness, which contains
the contribution of the extragalactic background as well. Therefore
we fitted parameters using a slightly modified version of Eq.~4:
\begin{equation}
  \rm {{N}\over{1\,mJy}} = C_0^{*} + C_1^{*}{\cdot}
  {\bigg\langle {{B(\lambda)-B_{CFIRB}(\lambda)}\over{1\,MJy\,sr^{-1}}} 
       \bigg\rangle}^{\eta^{*}}  
\end{equation}
\noindent 
$B_{CFIRB}$ is the surface brightness of the 
cosmic far-infrared
background at a specific wavelength \citep[c.f.][]{Pei99} and 
$B-B_{CFIRB}$\,=\,$B_{cirr}$ is the cirrus surface brightness. 
If $B_{cirr}$\,=\,0, then $N$\,=\,$C_0^{*}$, i.e. the confusion noise 
is purely due to cosmic infrared background fluctuations.
This indicates that $C_1^{*} \cdot \langle B_{cirr}\rangle^{\eta^{*}} $ is the
pure cirrus confusion noise component. We used these $C_1^{*}$ and 
$\eta^{*}$ values, together with the coefficients in Table~8 
for the creation of the all-sky cirrus confusion noise maps. 
The $C_0^{*}$, $C_1^{*}$ and ${\eta^{*}} $ parameters are listed in the
headers of the FITS files. }

%%%
\begin{figure}[h!!!]
\centerline{\hbox{\epsfig{file=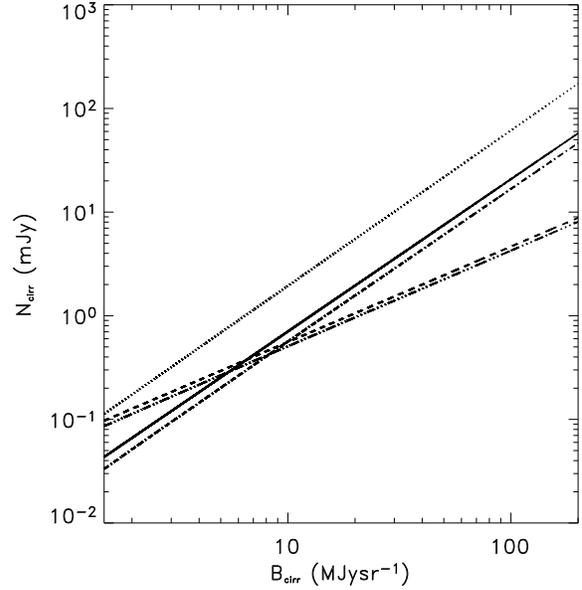, width=8.5cm}}}
\caption[]{Demonstration of the effects of the ISOPHOT--DIRBE 
 photometric transformation and variable spectral index $\alpha$ for 
 the Herschel/PACS 175\,$\mu$m filter cirrus 
 confusion noise predicions. 
 Solid line: no DIRBE--ISOPHOT transformation, constant $\alpha$;
 Dashed line: no DIRBE--ISOPHOT transformation, variable $\alpha$;
 Dash-dotted line: DIRBE--ISOPHOT conversion applied, constant $\alpha$;
 Dash-triple-dotted line: DIRBE--ISOPHOT conversion applied, variable $\alpha$.
 The dotted line shows the prediction according to \citet{Helou}.}
\label{fig:corrdemo}
\end{figure}
%%%

There are two other issues which 
have to be considered in the production of the confusion noise prediction maps: 
\begin{itemize}
\item[1.)] {\it Transformation between the COBE/DIRBE and ISOPHOT 
photometric systems}. The all-sky surface brightness maps are
in the DIRBE photometric system, while the coefficients
are derived in the ISOPHOT system. 
We used the latest available transformation coefficients
based on the comparison of DIRBE surface brightness values and 
ISOPHOT mini-map background fluxes obtained with PIA\,V10.0/CALG\,7.0
(Mo\'or, 2003, priv. comm.).
The transformation equation was:
  \begin{equation}
     B_{DIRBE}^{\lambda} = Gain{\times}B_{PHT}^{\lambda} + Offset
  \label{eq:dirbe}
  \end{equation}
where $B_{DIRBE}^{\lambda}$ and $B_{PHT}^{\lambda}$ are the COBE/DIRBE
and ISOPHOT surface brightness values, respectively.   
The transformation coefficients (Gain and Offset) for different filters are
summarized in Table~\ref{table:dirbe}.
%%%%
 \begin{table}
 \begin{center} 
 \begin{tabular}{rrr}
 \hline
 filter &   Offset        &   Gain \\
        &  (MJysr$^{-1}$) &         \\  \hline
   90\,$\mu$m & --1.65$\pm$0.06 & 1.00$\pm$0.01 \\  
  100\,$\mu$m &   0.09$\pm$0.09 & 0.85$\pm$0.02 \\
  120\,$\mu$m &   0.62$\pm$0.28 & 0.95$\pm$0.07 \\  
  150\,$\mu$m &   0.28$\pm$0.15 & 1.02$\pm$0.03 \\  
  170\,$\mu$m &   0.09$\pm$0.09 & 1.06$\pm$0.02 \\ 
  180\,$\mu$m & --0.50$\pm$0.32 & 1.15$\pm$0.09 \\  
  200\,$\mu$m &   0.95$\pm$0.17 & 0.97$\pm$0.03 \\ 
 \hline    
 \end{tabular}
 \end{center}
 \caption{Transformation coefficients between the ISOPHOT 
 and the DIRBE photometric systems, according to Eq.~\ref{eq:dirbe}.}
 \label{table:dirbe}
 \end{table}
%%%%

\item[2.)] {\it Constant or variable spectral index ($\alpha$)}.
As shown in Paper~II, the spectral index $\alpha$ depends on the
surface brightness of the observed fields and on the observational 
wavelength. The surface brightness dependence of $\alpha$ is approximated by:
%%%
\begin{equation}
\rm \alpha\,=\,A_1 \,\,{\times}\,\, 
log_{10} \bigg\langle {{B-B_{CFIRB}}\over{1\,MJy\,sr^{-1}}} \bigg\rangle
\,\,+A_0
\label{eq:sfbr-vs-alpha}
\end{equation}
%%%
\noindent where B is the average surface brightness of the field after 
the removal of the Zodiacal Light contribution and B$_{CFIRB}$ is the 
surface brightness of the cosmic far-infrared background. 
%\citep[see][for a review]{Hauser+Dwek}.  
The coefficients are: A$_0$\,=\,--1.67$\pm$0.47 and 
A$_1$\,=\,--1.57$\pm$0.38 for long wavelength filters (170/200\,$\mu$m)
of the C200 detector (see Paper~II). For shorter wavelength filters
(C100, 90 \& 100\,$\mu$m) this relation cannot be properly derived, 
therefore we do not use variable spectral indices for 
$\lambda$\,$<$\,120$\mu$m.

\end{itemize}

\noindent Due to these considerations two/four maps can be produced for each instrument:  
\begin{itemize}
\item[i)] No DIRBE -- ISOPHOT transformation, constant $\alpha$\,=\,--3 
\item[ii)] No DIRBE -- ISOPHOT transformation, \\ variable $\alpha$ 
	(only for $\lambda$\,$>$\,120$\mu$m)
\item[iii)] DIRBE -- ISOPHOT transformation applied, \\ constant $\alpha$\,=\,--3
\item[iv)] DIRBE -- ISOPHOT transformation applied, \\ variable $\alpha$
	(only for $\lambda$\,$>$\,120$\mu$m)
\end{itemize}
The detailed technical description of these all-sky fits files 
is given in Appendix~A.2.

The effect of these corrections is demonstrated in 
Fig.~\ref{fig:corrdemo} for the Herschel/PACS
175\,$\mu$m filter. The effect of the DIRBE--ISOPHOT transformation
applied or not is not very significant. In the contrary, the variable
$\alpha$ modifies the slope of the N$_{cirr}$--B$_{cirr}$ relation,
leading to a factor of $\sim$5 lower cirrus confusion noise at high 
surface brightness values.  

\begin{figure*}
\centerline{\epsfig{file=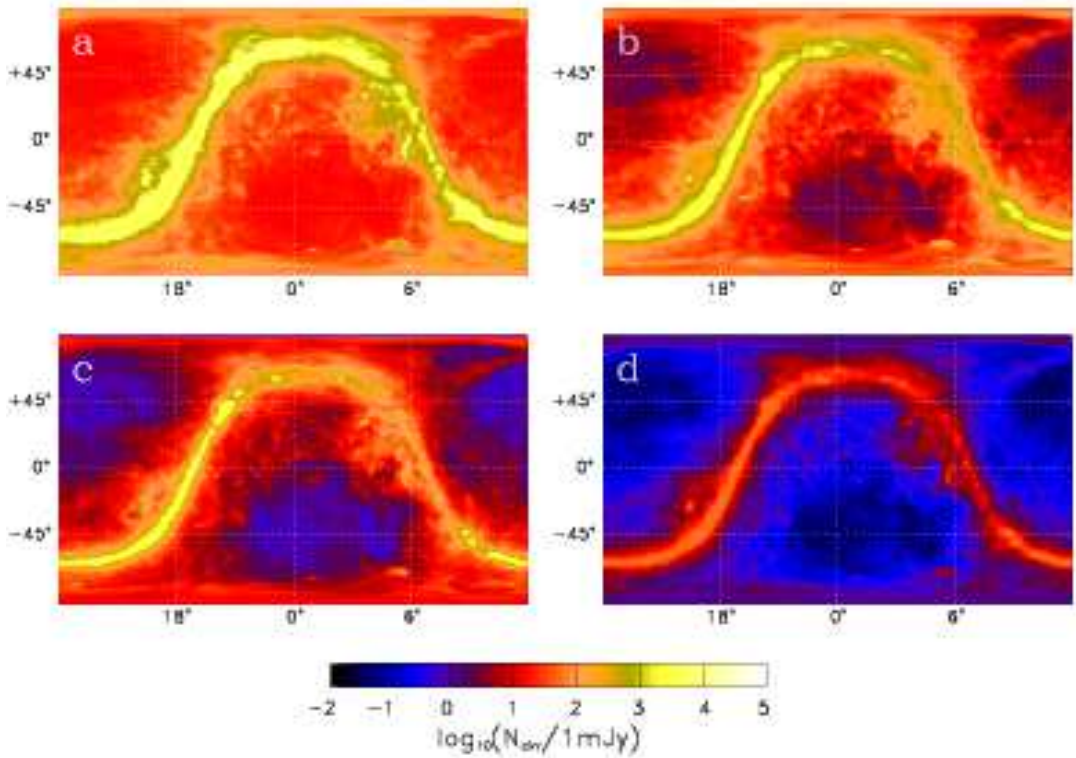, width=11.5cm, angle=90}}
\caption[]{Comparison of predicted cirrus confusion noise levels of four
different FIR instruments at their resolution limits. Confusion noise levels are 
displayed over the range as given by the colour-bar. All maps were constructed
with \emph{no} DIRBE--ISOPHOT conversion applied and with \emph{constant} $\alpha$
and are presented in an \emph{equatorial} coordinate system (J2000). 
({\bf a}) ISO/ISOPHOT C200 camera, 170\,$\mu$m;
({\bf b}) ASTRO-F/FIS 170\,$\mu$m;
({\bf c}) Spitzer/MIPS 160\,$\mu$m;
({\bf d}) Herschel/PACS 175\,$\mu$m. 
For a surface brightness of B\,$>$\,200\,MJy\,sr$^{-1}$
the cirrus confusion noise values given in this figure are \emph{lower limits} 
and represent the confusion noise value at B\,$=$\,200\,MJy\,sr$^{-1}$.}
\label{fig:common_predict_170}
\end{figure*}

An example comparing the capabilities of four instruments
(ISOPHOT~170\,$\mu$m, ASTRO-F/FIS~170\,$\mu$m, 
Spitzer/MIPS~160\,$\mu$m, Herschel/PACS~175\,$\mu$m) 
is given in Fig.~\ref{fig:common_predict_170}.
%%% 
%\subsubsection{Comparison of instrument capablilites}

%%
%%

%%%%%%%%%%%%%%%%%%%%
\section{Summary}

In this paper we present the results of a detailed investigation
of the dependence of the sky confusion noise 
and in particular the cirrus confusion noise on measurement configurations
for the long-wavelength ($\lambda$\,$\ge$\,90\,$\mu$m) observations with
ISOPHOT. Tables~2--7 together with Eq.~4 provide an easy tool 
to estimate confusion noise values. 
Based on these results and transformations for non-investigated filters
we constructed all-sky confusion noise maps
(stored as FITS files at: http://kisag.konkoly.hu/ISO/toolbox.html)
for all possible measurement configurations of the P3 100\,$\mu$m, 
C100 90, 100 and 105\,$\mu$m  and C200 120, 150, 170, 180 and 200\,$\mu$m
filters. These files can efficiently be used to estimate the confusion
noise at any specific sky position even if the background surface brightness
cannot be properly estimated from the measurement itself.  
From Version~7 onward the ISO Data Archive provides confusion noise 
values for individual ISOPHOT measurements, based on our numbers. 

The results of the ISOPHOT confusion noise analysis and the 
utilization of simulated fractal maps allowed us to
calculate cirrus confusion noise value ratios of ISOPHOT and 
other far-infrared spaceborn instruments. 
Using these values, all-sky maps with cirrus confusion noise estimates
for point sources were constructed. This was done for the resolution
limits of detectors of Spitzer/MIPS, ASTRO-F/FIS and Herschel/PACS. 
These maps can be used for the preparation of FIR observations with future 
space telescopes indicating sensitivity limits due to cirrus confusion noise.
However, for passively cooled large telescopes (like HERSCHEL) other noise 
components like the thermal telescope background 
\citep[see e.g.][]{Okumura} 
or the extragalactic background will play a more dominant role.   
%%%%%%%%%%%%%%%%%%%%%%%%%%%%%%%%%%%%%%%%%%%%%%%%%%%%%%%%%%%%%%%%%

\begin{acknowledgements}
\sloppy\sloppy
The development and operation of ISOPHOT were supported by MPIA and
funds from Deut\-sches Zentrum f\"ur Luft- und Raumfahrt 
(DLR). The ISOPHOT Data Center at MPIA is supported
by Deut\-sches Zentrum f\"ur Luft- und Raumfahrt e.V. (DLR) with
funds of Bundesministerium f\"ur Bildung und Forschung, 
grant~no.~50~QI~0201. 
Cs.K. acknowledges the support of the
ESA PRODEX programme (No.~14594/00/NL/SFe) and the Hungarian
Scientific Research Fund (OTKA, No.~T037508). 
We thank the referee for the useful comments and suggestions. 

\end{acknowledgements}
%%%%%%%%%%%%%%%%%%%%%%%%%%%%%%%%%%%%%%%%%%%%%%%%%%%%%%%%%%%%%%%%%%

%\newpage
%%%%%%%%%%%%%%%%%%%%%%%%%%%%%%%%%%%%%%%%%%%%%%%%%%%%%%%%%%%%%%%%%%

%%%%%%%%%%%%%%%%%%%%%%%%%%%%%%%%%%%%%%%%%%%%%%%%%%%%%%%%%%%%%%%%%%

%%%%%%%%%%%%%%%%%%%%%%%%%%%%%%%%%%%%%%%%%%%%%%%%%%%%%%%%%%%%%%%%%%


\begin{thebibliography}{99}
%\expandafter\ifx\csname natexlab\endcsname\relax\def\natexlab#1{#1}\fi

\bibitem[Calabretta \& Greisen(2002)]{Calabretta}
Calabretta, M. R., Greisen, E. W., 2002, A\&A 395, 1077
%
\bibitem[Dole et al.(2003)]{Dole}
Dole, H., Lagache, G., Puget, J.-L., 2003, ApJ 585, 617
%
\bibitem[Elmegreen(1997)]{Elmegreen}
Elmegreen, B.G., 1997, ApJ 477, 196
%
\bibitem[Gabriel et al.(1997)]{Gabriel97}
Gabriel, C., Acosta-Pulido, J., Heinrichsen, I., Morris, H., Tai, W.-M., 
1997, The ISOPHOT Interactive Analysis PIA, a Calibration and 
Scientific Analysis Tool, in: ADASS VI., ASP Conf. Ser. Vol. 125,
G. Hunt and H.E. Payne (eds.), p. 108
%
\bibitem[Gautier et al.(1992)]{Gautier}
Gautier\,III, T.N., Boulanger, F., P\'erault, M., {Puget} J.L., 1992, 
AJ 103, 1313 
%
\bibitem[Hanish et al.(2001)]{Hanish}
Hanisch, R.J., Farris, A., Greisen, E.W., et al., 2001, A\&A 376, 359
%
\bibitem[Hauser \& Dwek(2001)]{Hauser+Dwek}
Hauser, M.G., Dwek, E., 2001, ARA\&A 39, 249
%
\bibitem[Helou \& Beichman(1990)]{Helou}
{Helou}, G., {Beichman}, C.A., 1999, The confusion limits to the sensitivity 
of submillimeter telescopes, in: From Ground-Based to Space-Borne Sub-mm 
Astronomy, Proc. of the 29th Li\`ege International Astrophysical Coll., 
ESA Publ., p. 117. 
%
\bibitem[Herbstmeier et al.(1998)]{Herbstmeier}
Herbstmeier, U., \'Abrah\'am, P., Lemke, D., et al., 1998, A{\&}A 332, 739 
%
\bibitem[Ingalls et al.(2004)]{Ingalls}
{  Ingalls, J.G., Miville-Desch\^enes, M.-A., Reach, W.T., et al., 
2004, ApJS, accepted {\tt [astro-ph/0406237]}}
%
\bibitem[Jeong et al.(2003)]{Jeong}
Jeong, W.-S., Pak, S., Lee, H.M., et al., 2003, PASJ 55, 717
%
\bibitem[Kessler et al.(1996)]{Kessler96}
Kessler, M.F., Steinz. J.A., Anderegg, M.F., et al., 1996, A{\&}A 315, L27
%
\bibitem[Kessler et al.(2003)]{Kessler2003}
Kessler, M.F., M\"uller, Th.G., Leech, K., et al., 
2003, The ISO Handbook Vol. I.: ISO -- Mission \& Satellite Overview, 
Version 2.0, ESA SP-1262, European Space Agency  
%
\bibitem[Kiss et al.(2001)]{Kiss2001}
Kiss, Cs., \'Abrah\'am, P., Klaas, U., Juvela, M., Lemke, D., 2001,
A\&A 379, 1611 (Paper~I)
%
\bibitem[Kiss et al.(2003)]{Kiss2003}
Kiss, Cs., \'Abrah\'am, P., Klaas, U., et al., 2003, A\&A 399, 177  (Paper~II)
%
\bibitem[Laureijs et al.(2003)]{Laureijs2003}
Laureijs, R.J., Klaas, U., Richards, P.J., Schulz, B., \'Abrah\'am, P.,
2003, The ISO Handbook Vol. IV.: PHT -- The Imaging Photo-Polarimeter,
Version 2.0.1, ESA SP-1262, European Space Agency  
%
\bibitem[Lemke et al.(1996)]{Lemke96}
Lemke, D., Klaas, U., Abolins, J., et al., 1996, A{\&}A 315, L64
%
\bibitem[Miville-Desch\^enes et al.(2002)]{MD2002}
{  Miville-Desch\^enes, M.-A., Lagache, G., Puget, J.-L., 2002,
A\&A 393, 749}
%
\bibitem[Miville-Desch\^enes et al.(2003)]{MD2003}
{  Miville-Desch\^enes, Joncas, G., Falgarone, E., Boulanger, F., 
2003, A\&A 411, 109}
%
\bibitem[Negrello et al.(2004)]{Negrello}
Negrello, M., Magliocchetti, M., Moscardini, L., 2004, 
[astro-ph/0401199]
%
\bibitem[Okumura(2001)]{Okumura}
Okumura, K., 2001, Chopping noise due to the primary 
mirror of HERSCHEL (Herschel Space Telescope internal report), 
SAp-PACS-KO-0087-02
%
\bibitem[Okumura \& Longval(2001)]{Okumura+Longval}
Okumura, K., Longval, Y., 2001, HERSCHEL telescope PSF modeling 
(Herschel Space Telescope internal report), SAp-FIRST-KO-0094-02
%
\bibitem[Pei et al.(1999)]{Pei99}
Pei, Y.C., Fall, M.S., Hauser, M.G., 1999, ApJ 522, 604
%
\bibitem[Press et al.(1992)]{NumRec}
Press, W.H., Teukolsky, S.A., Vetterling, W.T., Flannery, B.P., 
1992, Numerical Recipes in Fortran~77: The Art of Scientific Computing, 
2$^{nd}$ edition, Cambridge University Press
%
%\bibitem[1998]{Stutzki}
%Stutzki J., Bensch F., Heithausen A., Ossenkopf V., Zielinsky M., 
%1998, A\&A 336, 697
\end{thebibliography}
\end{document}